\documentclass[preprint,twocolumn]{aastex6}
\usepackage{subfigure}
\usepackage{color}
\usepackage{amsmath}
\usepackage{mathrsfs}
\usepackage{multirow}
\usepackage{soul}
\usepackage{amssymb}
\usepackage{hyperref}
\def\be{\begin{equation}}
\def\ee{\end{equation}}
\def\bd{\begin{displaymath}}
\def\ed{\end{displaymath}}
\def\ba{\begin{aligned}}
\def\ea{\end{aligned}}
\def\nms{\mathsurround=0pt}
\def\oversim#1#2{\lower 4pt\vbox{\baselineskip 0pt \lineskip 1pt
    \ialign{$\nms#1\hfil##\hfil$\crcr#2\crcr\sim\crcr}}}
\def\ga{\mathrel{\mathpalette\oversim>}}
\def\la{\mathrel{\mathpalette\oversim<}}

\def\bh{M_{\bullet}}

\def\msun{M_{\odot}}
\def\AU{{\rm AU}}

\def\pl{\partial}
\begin{document}

\title{Gravitational-wave merging events from the dynamics of stellar mass
binary black holes around the massive black hole in a galactic nucleus}

\author{FUPENG ZHANG$^{1,2}$, Lijing Shao$^3$, Weishan Zhu$^2$}
\affil{$^1$\,School of Physics and Electronic Engineering, Guangzhou
University, 510006 Guangzhou, China, zhangfupeng@gzhu.edu.cn\\
$^2$\,School of Physics and Astronomy, Sun Yat-Sen University, Guangzhou
510275, China\\
$^3$\,Kavli Institute for Astronomy and Astrophysics, Peking University,
Beijing 100871, China }

\begin{abstract}
We study the dynamical evolution of the stellar mass binary black holes (BBHs)
in a galactic nucleus that contains a massive black hole
(MBH). For a comprehensive study of their merging events, we consider
simultaneously the non-resonant and resonant
relaxations of the BBHs, the binary-single encounters
of the BBHs with the field stars, the Kozai-Lidov (KL) oscillation and the
close encounters between the BBHs and the central MBH, which usually lead
to binaries' tidal disruptions.  As the BBHs are usually
heavier than the background stars, they sink to the center by
mass segregation, making the KL oscillation an important
effect in merging BBHs.  The binary-single encounters can not only lead to
softening and ionization of the BBHs, it can also
make them hardening, that increases the merging rates significantly. The
mergers of BBHs are mainly contributed by galaxies
containing MBHs less massive than $10^8\msun$ and the total event rates
are likely in orders of $1$--$10$ Gpc$^{-3}$ yr$^{-1}$,
depending on the detailed assumptions of the
nucleus clusters. About $3-10\%$ of these BBH mergers are with eccentricity
$\ge 0.01$ when their gravitational wave oscillating frequencies enter the LIGO band ($10$\,Hz).
Our results show that merging the BBHs within galactic nuclei
can be an important source of the merging events
detected by the Advanced LIGO/Virgo detectors,
and they can be distinguished
from BBH mergers from the galactic fields and globular clusters when enough events
are accumulated.
\end{abstract}


\section{Introduction}

The first gravitational-wave (GW) event GW150914
has been discovered by the Advanced
LIGO~\citep{Abbott16a}. It is caused by the merger of a
stellar mass binary black hole (BBH) with component
masses given by $\sim36\msun$ and $\sim29\msun$. Later on, a number of
stellar-mass BBH mergers were reported
~\citep[for the latest results see][]{Abbott18},
e.g., GW151226~\citep{Abbott16c},
GW170104~\citep{Abbott17a}, GW170608~\citep{Abbott17b},
and GW170814~\citep{Abbott17c}. Currently, the locations of these merging events
and corresponding mechanisms remain largely uncertain. The proposed
explanations include (1) merging BBHs in the isolated
environments~\citep[e.g.,][]{Dominik15,Belczynski16}; (2) merging BBHs in the
globular clusters~\citep[e.g.,][]{Morscher15,Rodriguez15, Rodriguez16,Fragione18b}; (3)
merging BBHs in the galactic nuclei. For example, the
GW capture of two encountering black
holes~\citep{Oleary09}, and the merging of BBHs due to
the Kozai-Lidov (KL) effect~\citep[e.g.,][]{Antonini10, Antonini12,VanLandingham16,Petrovich17}.
Such merging
events can be possible without the presence of
a central massive black hole (MBH)~\citep{Miller09, Antonini16}.
So far, it is still challenging to distinguish these
different mechanisms from the observations, as localizations
are poor, and that electromagnetic counterparts
for these events have not been found convincingly.

A galactic nucleus that harbors a MBH in the center is a very
complex system. Both the inner (orbit of the BBH
circling with each other) and outer (orbit of the BBH circling
the central MBH) orbits within such environments can be
affected by a number of different dynamical effects: (1) The dynamical evolution
of the outer orbit of the BBHs, due to the non-resonant (NR, or two-body
relaxation) or resonant relaxation (RR) effects~\citep{RT96}; (2) The
encounters between the BBHs and field stars (or any stellar remnants). Such
processes can change the inner orbit of the BBHs,
and lead to softening or hardening of the binary, or even exchange of
the binary components with the incoming
objects~\citep[e.g.,][]{Heggie93,Sigurdsson93,McMillan96,Downing10,Samsing14};
(3) A MBH and a BBH form a natural hierarchical
triple system, where the KL effects will lead to
the orbital oscillations~\citep[e.g.,][]{Kozai62, Naoz13, Naoz16}. In some
cases the eccentricity of the BBH can be excited
to an extremely high value such that the BBH will merge eventually due to the
GW radiation~\citep[e.g.][]{Antonini12,Hoang18}; (4) The tidal disruption of
the BBHs by the central MBH. The strong tidal force of the central MBH will
likely disrupt a BBH that is very close to it. Along the path that
leads to the merging of the BBHs, all of the above dynamical
processes may affect the evolution of the inner and outer orbits. Thus, to obtain
a realistic estimation of the merging
rates, all of the above dynamical processes should be simultaneously
considered. However, they have not been well combined in previous
studies yet.

We notice that the BBHs in a galactic nucleus are most likely soft binaries~\citep{Hopman09} and
to be merged within the galactic nucleus, different from those BBHs in
globular clusters, which are mostly hard binaries and likely to be
merged after they are ejected from the cluster~\citep{Rodriguez16}. Some
dynamical processes, e.g., KL effects,
RR, tidal disruptions of BBHs by the central MBH,
are unique, requiring the existence of a MBH, and they
happen only in the galactic nuclei.
Thus, to distinguish the mergers of stellar mass BBHs that take place in galactic nuclei and those resulting from globular clusters, it is important for us to trace all of the dynamical effects in
a galactic nucleus mentioned above.

Here in this work, we aim to build a comprehensive framework to study the
evolution of stellar mass BBHs in a galactic nucleus, for both the inner and the outer orbits,
by using Monte-Carlo methods that include all of the above effects.
Such
a framework can provide a more detailed estimation of
how the merging rate of BBHs is affected by individual, or combinations of
these effects.
To investigate the importance of the mergers of BBHs in
 galactic nuclei for the Advanced
LIGO/Virgo detectors, we can compare our predicted merging signals and merging rates in
the local universe with  observations. Our investigation can not
only improve the understandings of the  channels of BBHs merging in a
galactic nucleus, but also the dynamics of stars and binaries around the MBH.

The paper is organized as follows. We describe the simulation details of
various dynamical processes in Section~\ref{sec:method}. In
Section~\ref{sec:simulation} we perform a number of Monte-Carlo simulations to
show the effects of various dynamical processes on the merging
and evolution of the BBHs.  In Section~\ref{sec:event_rate} we provide
estimations of the merging rates of stellar mass BBHs in a galactic nucleus in
the nearby universe. Section~\ref{sec:simulation} concentrates
  mostly on the details of the BBH's dynamical evolution in a galactic nucleus, thus
readers who are more interested in the rate and properties of the
merging events of BBHs in the local universe, can skip Section~\ref{sec:simulation} and
turn to Section~\ref{sec:event_rate} directly.
Discussion and conclusion are provided in
Sections~\ref{sec:discussion} and~\ref{sec:conclusion}, respectively.
For clarity, some notations of variables that are frequently used in
this paper are summarized in Table~\ref{tab:symbols}.

\begin{table*}
\caption{Notation of Some Symbols}
\centering
\begin{tabular}{l|p{7cm}|l|p{7cm}}
\hline
Symbol & Description & Symbol & Description\\
\hline  
$E$, $J$ & The energy and angular momentum of the orbit circling around a MBH   & $\bh$      & The mass of the central MBH\\
$m_A$, $m_B$ & The masses of the two components of a BBH             & $\sigma_h$ & The velocity dispersion of the galactic nucleus \\
$m_{\rm BBH}$ & The total mass of a BBH                               & $r_h$ & The influence radius of a MBH \\
$m_T$ & The total mass of the BBH and the incoming star in the binary-single encounter & $n_\star$ & The number density of field stars \\
$a_1$, $e_1$ & The SMA and eccentricity of the inner orbit of a BBH          & $r_p$ & The pericenter of the outer orbit of a BBH\\
$a_2$, $e_2$ & The SMA and eccentricity of the outer orbit of a BBH & $r_t$ & The tidal radius of the BBH around a MBH\\
$P_1$, $P_2$ & The orbital period of the inner and outer orbits of a BBH & $g_1$, $g_2$ & The argument of periapsis of the inner and outer orbits of a BBH\\
$\Omega_1$, $\Omega_2$ & The ascending node of the inner and outer orbits of a BBH
& $I_1$, $I_2$ & The orbital inclination of the inner and outer orbits of a BBH \\
$\alpha_{\rm BBH}$ & The power law index of the density profile of BBHs  & $\alpha_\star$ & The power law index of the density profile of field stars \\
$x$ & $x=E/E_0$, the normalized energy of the orbit circling around a MBH.  Here $E_0=-G\bh/r_h$ & $v_c$ & Critical velocity that defines the hard and soft binary regions of binary-single encounters \\
$g_{\rm BBH}(x)$ &  Dimensionless distribution function of the
energy of the BBHs' outer orbit & $g_\star(x)$ & Dimensionless distribution function of the field stars' orbital energy\\
$m_\star$ & The mass of field stars                                &  $p$, $b$ & The closest distance and the impact parameter between a BBH and an incoming star in a binary-single encounter event\\
$T_{\rm GR}$ & The GR procession timescale of the inner orbit of a BBH & $T_{\rm GW}$ & The GW orbital decay timescale of the inner orbit of a BBH \\
$\mathscr{R}$ & The normalized GW event rate
&$e_{10\,\rm Hz}$ & The eccentricity of the BBH when its peak
frequency of GW radiation reaches to $10$\,Hz\\
\hline
\end{tabular}
\label{tab:symbols}
\end{table*}

\section{The Method}
~\label{sec:method}
\begin{figure*}
\center
\includegraphics[scale=0.7]{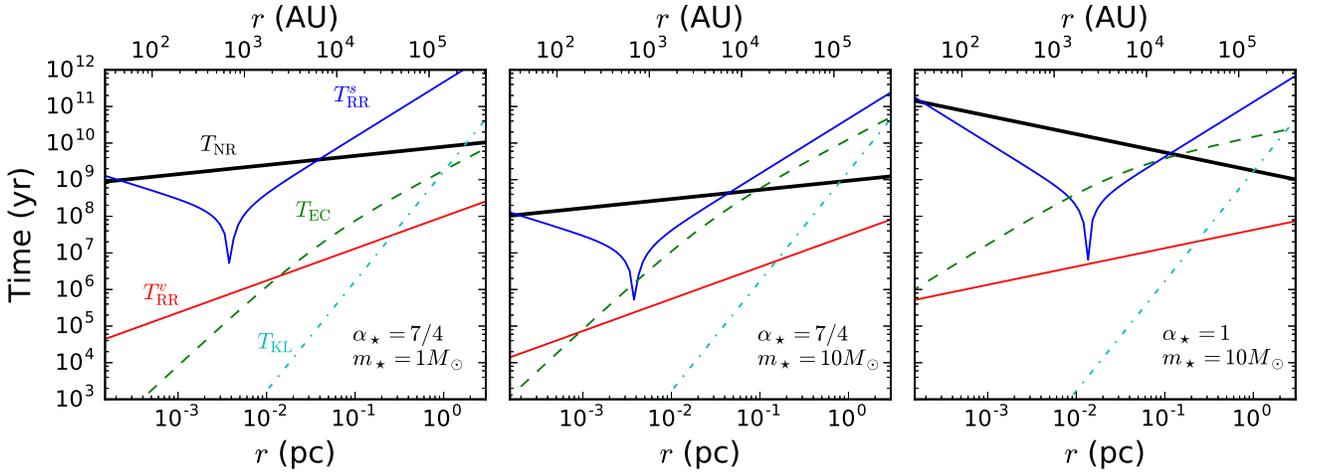}
    \caption{Timescales of different dynamical processes around a MBH
    with mass $\bh = 4\times10^6\msun$.  The mass of
    the binary is given by $m_A=m_B=10\msun$. $r$ is the distance from the MBH.
    We assume $e_2=0.6$ for the eccentricity of the outer orbit. $T_{\rm NR}$,
    $T_{\rm RR}^{s}$ and $T_{\rm RR}^{v}$ are the two body relaxation timescale,
    scalar RR timescale and vector RR timescale,
    respectively. $T_{\rm EC}=R_{\rm EC}^{-1}$ is the time required to have a
    collision between a BBH and a background star, here $R_{\rm EC}$ is given
    by Equation~\ref{eq:rbi}.  In all panels we have set $p=3a_1$.  $T_{\rm
    KL}$ is the KL oscillation period for a BBH-MBH triple system.  Different
    panels show the timescales when $\alpha_\star$ (the density profile of the
    field stars) and $m_\star$ (the mass of the background stars)
    are different, as labeled in the panels.
}
\label{fig:ts}
\end{figure*}

We consider a nucleus cluster that consists of field stars with the
same mass $m_\star$, and contains a MBH in the center with mass $\bh$. Then the
MBH dominates the dynamical evolutions of the cluster within the gravitational
influence radius $r_h$, given by~\citep{Hopman06}
\be
r_h=\frac{G\bh}{\sigma_h^2}=2.31~{\rm
pc}\left(\frac{\bh}{4\times10^6~\msun}\right)^{1/2},
\ee
where $\sigma_h$ is the velocity dispersion of the galaxy, and we have
used the well-established $\bh$-$\sigma$ relation.

The enclosed mass within $r_h$ is about the mass of the MBH, i.e., $M(<r_h)\sim \bh$. For an isotropic cluster, the number density of the field stars is given
by $n_\star(r)=n_0(r/r_h)^{-\alpha_\star}$, where $r$ is the distance from the MBH and
$\alpha_\star$ is the index of the density profile, $n_0\simeq (3-\alpha_\star)\bh/(4\pi
m_\star r_h^3)$ is the number density at distance $r_h$.

We assume that the BBHs are initially placed at distance $\sim r_h$ from the MBH
and migrate into the cluster due to dynamical processes (e.g.,
diffusion and mass segregation by two body scatterings). We assume that the
distribution of the field stars is initially in an equilibrium state, and that the
existence of the BBHs does not affect the distributions of the field
stars, as the total number and mass of BBHs should be always much smaller than
that of the field stars.

The dynamical evolutions of the BBH within $r<r_h$ are significantly different
from the outer parts.  Both the outer and inner orbits of the BBH can
be affected by various dynamical processes when they migrate into the inner
regions of the cluster. Denote $a_1$, $e_1$,
$\Omega_1$, $I_1$ and $g_1$ ($a_2$, $e_2$, $\Omega_2$, $I_2$ and $g_2$)
as the inner (outer) semimajor axis (SMA), eccentricity,
ascending node, inclination and the argument of periasis of the BBH's orbit,
respectively.
The outer orbit of the BBH is evolved mainly by
the NR and RR.  If $e_2$ is very high, the BBH may approach to $r_t$, i.e., the
tidal radius of the MBH (see Eq. \ref{eq:rt}), such that its inner orbit is strongly perturbed, or
likely to be disrupted after the encounter with the MBH. It is possible
that the BBHs experience multiple encounters with the field stars, which can
lead to cumulative modifications of $a_1$ and $e_1$, exchange of the binary
components with the incoming stars, ionization of the BBHs, or merging of the
BBHs by the GW orbital decay. When the BBH is slightly away from
the tidal radius, its $e_1$ may be excited to
a high value by KL oscillation, such that the
BBH quickly merges and triggers a GW event. During
all these processes, the orbital orientations (described by $\Omega_1$,
$I_1$ or $\Omega_2$, $I_2$) or other elements (e.g., $g_1$, $g_2$) of both the
inner and outer orbits of BBHs can be cumulatively changed,
affecting the evolutions and the merging of individual BBHs.
As all of these processes may be important for the understandings of the GW merging
events due to BBHs in galactic nuclei, here, our numerical method
considers all of the above dynamical processes.  The dynamical
timescales for these processes for a Milky Way-like galaxy can be found in
Figure~\ref{fig:ts}. The details of the processes are described in the
following subsections.

\subsection{Dynamics of the outer orbit}
\label{subsec:dy_outer}

Denote $E=-G\bh/a_2$ and $J=\sqrt{G\bh a_2(1-e_2^2)}$ as the energy and angular
momentum of the outer orbit of a BBH.
Both $E$ and $J$ are evolved under the NR. $J$ of the
BBH is additionally affected by RR. The NR is due to the weak
gravitational interactions between a BBH and field stars, where
we treat the BBH as a point mass particle.
The NR (or two-body relaxation) timescale is given by~\citep{Binney87}
\be\ba
T_{\rm NR}&=\frac{0.34\sigma^3}{(Gm_\star)^2n_\star ln\Lambda},\\
\label{eq:tnr}
\ea\ee
where $ \Lambda\simeq \bh/m_\star$ \citep{BW76}.
Here $\sigma\propto r^{-1/2}$, thus $T_{\rm NR}\propto
r^{\alpha_\star-3/2}$.  The time evolutions of the diffusion process in the $E-J$
space due to two-body relaxation can be described by the Fokker-Planck
equation, and simulated by a Monte-Carlo scheme according to their
orbit-averaged diffusion coefficients, i.e., $D_{EE}$, $D_{E}$, $D_{JJ}$,
$D_J$ and $D_{EJ}$ ~\citep{LS77}. Here $D_{EE}$ ($D_{JJ}$) and $D_{E}$ ($D_J$)
describe the orbit-averaged scatterings of the energy (angular momentum) and its drift,
respectively.  $D_{EJ}$ describes the correlations between the
scatterings of energy and angular momentum.  The details of the formalisms
 are given in Appendix~\ref{apx:two_body}.  We adopt a
Monte-Carlo scheme similar to~\citet{SM78} and~\citet{Baror16} to calculate the
two-body relaxation in $E-J$ of BBHs due to the field stars. More details will be explained in Section~\ref{subsec:MC_schemes}.

In the inner part of the cluster (but not too close to the MBH, so the general
relativistic precession is still not significant), RR becomes important, and the resonant
torques between the orbits of the field stars can change quickly the orbital
angular momentum of the BBHs. The RR changes both the amplitude (scalar
RR) and the direction of the angular momentum (vector RR).  The timescale of
scalar RR is given by~\citep{Hopman06}
\be
T_{\rm RR}^s=\frac{A_{\rm
RR}^s}{t_\omega}\left(\frac{\bh}{m_\star}\right)^2\frac{P^2(a_2)}{N_\star(<a_2)},
\ee
where $A_{\rm RR}^s\simeq3.56$~\citep{RT96}. $t_{\omega}=2\pi/\nu_p$ is the
timescale of orbital precession and $\nu_p$ is given by Equation~\ref{eq:nup}.
$P$ is the orbital period and $N_\star(<a_2)$ is the number of field stars
within
distance of $a_2$.
The scalar RR can be suppressed due to the rapid general relativistic
orbital precession below the locus called ``Schwarzschild barrier'' (SB)
~\citep{Merritt11,Antonini13}, where the orbital precession frequency equals to the coherence frequency, i.e.,
$\nu_p=2\pi/T_c(a_2)$~\citep{Baror16}. Here $T_c(a_2)$ is the coherence timescale
given by Equation~\ref{eq:tca2}.

The timescale of vector RR is given by~\citep{Hopman06}
\be
T_{\rm RR}^v=2A_{\rm
RR}^v\left(\frac{\bh}{m_\star}\right)\frac{P(a_2)}{N_\star^{1/2}(<a_2)},
\label{eq:trrv}
\ee
where $A_{\rm RR}^v\simeq0.31$~\citep{RT96}.  By a method similar
to~\citet{Baror16}, we can consider the scalar RR by calculating the diffusion
coefficients in angular momentum $D_{JJ}^{RR}$ and $D_{J}^{RR}$. For vector RR,
we simply consider it as a diffusion process. The details of the scalar and
vector RR are shown in Appendix~\ref{apx:resonant}.

\subsection{Tidal disruption of BBHs}
\label{subsec:tidal_disruption}

During the evolution, if the BBH approaches too close to the MBH, they
will likely be disrupted by the tidal force of the MBH.  The tidal radius of
binary stars is given by
\be
r_t=\left(\frac{3\bh}{m_{\rm BBH}}\right)^{1/3}a_1,
\label{eq:rt}
\ee
where $m_{\rm BBH}=m_A+m_B$ is the total mass of the binary.  The loss cone of the
angular momentum is $J_{\rm lc}^2\simeq2\bh r_t$.  The empty loss cone region
requires that the change of angular momentum per orbital period of the
BBH is smaller than the size of lose cone, i.e., $dJ<J_{\rm lc}$ and
the full loss cone region requires that $dJ>J_{\rm lc}$. In
full lose cone region, the binary can jump in and out of the lose cone multiple
times before it encounters with the MBH. In empty loss cone
region, the binary takes multiple periods to move into the lose cone and may
have encountered with the MBH multiple times. As the probability of
tidal disruption is a function of distance from the MBH (e.g., in the case of
binary stars~\citep{Zhang10}). The BBH may experience multiple
encounters with the MBH before the tidal disruption. There
are also chances that the BBH moves
away from the lose cone after the multiple encounters with the MBH.
\citet{Zhang10} have showed that the multiple encounters can change both the SMA
and the eccentricity of the binary, which thus may lead
to merging events.

To consider accurately such effects, we use three-body numerical integrations
to calculate the outcome of each encounters between the BBH and the
MBH. Considering that the MBH can impose strong impacts on the inner orbit of
the BBH even before it reaches the loss cone, we set up the
three body integration when $a_2(1-e_2)<3r_t$. At the beginning of
these explicit three-body simulations, initially the binary is placed at
$10^4a_1$ (or at apocenter of the outer orbit if $a_2$ is smaller).
After the encounter, the change of the inner orbit of the binary at distance
$10^4a_1$ from the MBH is recorded for next encounters.
Such a process repeats till either the binary is disrupted, or merged due to
the GW emission,
or moving out of the loss cone.

\subsection{Kozai-Lidov Effect and the Gravitational wave orbital decay}
\label{subsec:kl_gw}

\begin{figure*}
\center
\includegraphics[scale=0.75]{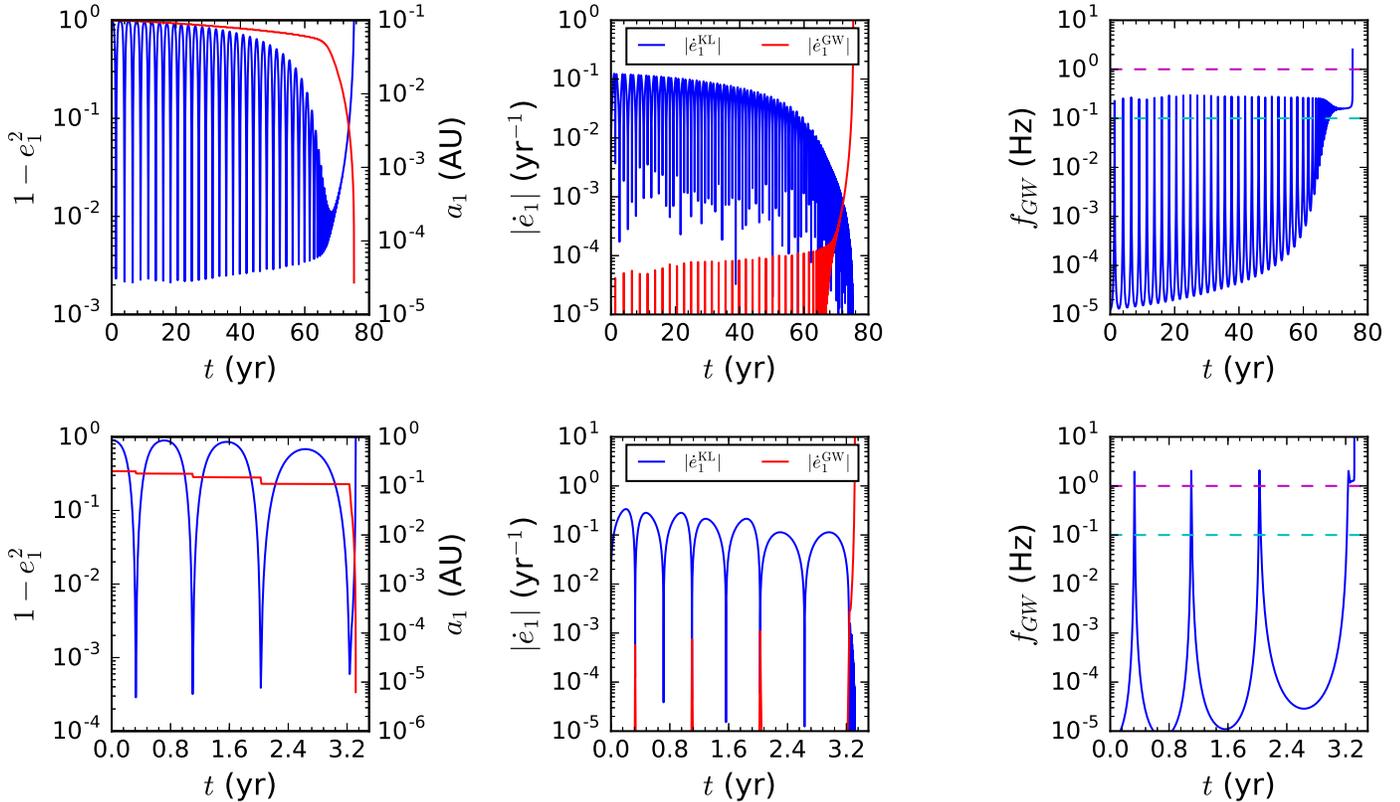}
\caption{Left panels: The time evolution of $1-e_1^2$ (solid blue lines) and
  $a_1$ (solid red lines) of the BBHs before they merge.  Middle panels: The
  time evolution of $|\dot e^{\rm GW}_1|$ (solid blue lines) and $|\dot e_1^{\rm
  KL}|$ (solid red lines), which are given by Equation~\ref{eq:dae1_gw} and
  from~\citet{Naoz13}, respectively. Right panels: The time evolution of the
  peak of GW frequency, given by Equation~\ref{eq:fgwob}.  The magenta dashed
  line shows $f_{\rm GW}=1\,$Hz, above which the GW can be
  observed by the ground-based GW detectors. The cyan dashed line
  shows $f_{\rm GW}=0.1$\,Hz. The upper panels are results from a BBH
  assuming $a_1=0.1\AU$, $e_1=0.1$, $m_A=10 \, \msun$, $m_B=20\,\msun$. The peak frequency
  $f_{\rm GW}$ is larger than $1$Hz {\it after} the GW dominates the evolution of the BBH.
  For bottom panels, the assumed
  BBH is similar but with $a_1=0.2\AU$ and $e_1=0.3$.  Different from the upper
  panels, the peak frequency is larger than $1$Hz {\it before}
  the GW dominates the evolution of the BBH. For all panels the outer
  orbit of the BBH is given by $a_2=50 \, \AU$ and $e_2=0.3$.
}
\label{fig:klgw}
\end{figure*}

The BBH-MBH system forms a natural triple system and the tidal
force of the MBH can trigger KL oscillation on the inner and outer orbits of
the BBH. The KL effect will increase the eccentricity of
the inner orbit of the binary under some preferred orbital configurations
~\citep[e.g.][]{Naoz13,Naoz16}.  The period of KL oscillation is given
by~\citep{Kiseleva98}
\be\ba
T_K&=\frac{2P_2^2}{3\pi P_1}(1-e_2^2)^{3/2}\frac{m_A+m_B+\bh}{\bh}\\
\simeq &\frac{2}{3^{1/3}\pi} \left(\frac{r_p}{r_t}\right)^{3/2}P_2.
\ea\ee
The GR procession of the inner orbit of the binary is given by
\be
T_{\rm GR}=\frac{2\pi a_1^{5/2}c^2(1-e_1^2)}{3 G^{3/2} (m_A+m_B)^{3/2}}.
\ee
If $T_{\rm GR}<T_K$, or,
\be\ba
r_p\la 0.58 \left(\frac{T_{\rm GR}}{P_2}\right)^{2/3}r_t,
\ea\ee
the KL effect will be suppressed~\citep[e.g.,][]{Ford00, Naoz13}. In
most cases, the KL effect is only significant when the pericenter of the outer
orbit of the BBH is close to the tidal radius of the MBH.  In this work, we
consider the KL effect only when $r_p$ falls in the range of $3r_t$ to ${\rm min} [20, 0.58(T_{\rm GR}/P_2)^{2/3}]r_t$.  When $r_p<3r_t$ we use the explicit three-body
integration to calculate the interactions between the BBHs and
the MBH (See Section~\ref{subsec:tidal_disruption}).

When the eccentricity of the inner orbit, $e_1$, is excited to
a high value due to KL effect (or other effects), the GW
radiation becomes important and the BBH is likely
to be merged in a very short timescale. The GW orbital decay timescale is
given by ~\citep{Peters64}
\be
T_{\rm GW}=\frac{5}{64}\frac{c^5a^4}{m_Am_B(m_A+m_B)G^3}
\frac{(1-e_{1}^2)^{7/2}}{1+\frac{73}{24}e_{1}^2+\frac{37}{96}e_{1}^4}.
\ee
The evolution of the SMA and eccentricity of the inner binary by GW radiation
is given by~\citep{Peters64}
\be\ba
{\dot a_1}^{\rm GW}&=-\frac{64}{5}\frac{m_Am_B(m_A+m_B)G^3}{c^5a_1^{3}}
\frac{1+\frac{73}{24}e_{1}^2+\frac{37}{96}e_{1}^4}{(1-e_{1}^2)^{7/2}},\\
{\dot e_1}^{\rm GW}&=-\frac{304}{15}e_{1}\frac{m_Am_B (m_A+m_B) G^3}{c^5a_1^{4}}
\frac{1+\frac{121}{304}e_{1}^2}{(1-e_{1}^2)^{5/2}}.
\label{eq:dae1_gw}
\ea\ee
Here the dot means the derivative respect to time $t$. When the
eccentricity is high, the argument of pericenter (denoted as $g_1$) of the
BBH's inner orbit can precess significantly. For simplicity, we
consider only the post-Newtonian GR effect, i.e.,
\be
{\dot g_1}^{\rm PN}=\frac{3 (m_A+m_B)^{3/2} G^{3/2}}{c^2a_1^{5/2}(1-e_1^2)}.
\ee

We consider the evolution of the inner and outer orbits of the BBHs
under the above two effects and the KL oscillations. For KL effect, we adopt
the time evolution formalism in~\citet{Naoz13}. The equations of motion in
$a_1$, $e_1$ and $g_1$ now become
\be\ba
\dot a_1={\dot a_1}^{\rm GW},~&~
\dot e_1={\dot e_1}^{\rm GW}+{\dot e_1}^{\rm KL},~&~
\dot g_1={\dot g_1}^{\rm PN}+{\dot g_1}^{\rm KL}.
\ea
\ee
Here ${\dot e_1}^{\rm KL}$, ${\dot g_1}^{\rm KL}$ can be found
in~\citet{Naoz13}.  For the KL evolution of the other
orbital elements of BBHs, see~\citet{Naoz13}\footnote{
We have noticed that~\citet{Naoz13} does not consider the gravitational
radiation loss of the total angular momentum $H$ (see~\citet{Blaes02}).
Nevertheless, it should not affect much of our simulation results,
as such radiation loss on $H$ should be only important when
the gravitational wave orbital decay dominates the motion, and could be ignored
during most time of the KL oscillation.
}. Recent studies show that these analytical formula capture only the
secular terms of the evolution, while the non-secular terms may be important
for a BBH-MBH triple system~\citep{Grishin18}. The non-secular terms can enhance
the maximum eccentricity in oscillations and increase the merging rates
by up to an order of magnitude~\citep{Fragione18a}.
Thus, our merging rates due to KL effect may have been underestimated.

It would be interesting to investigate the eccentricity and the SMA when the
peak of GW frequency enters to the observing band of
ground-based GW detectors, i.e., $f_{\rm GW} \ga 1$ Hz.
We discribe the method in Appendix~\ref{apx:GWfreq}, that mainly bases on the
Equations~\ref{eq:fgwob} and~\ref{eq:gwae}.
Figure~\ref{fig:klgw} illustrates the two possible cases of merging a BBH.
The first one is shown by the top panels, i.e., when the
 GW frequency approaches to $f_{\rm GW}=1$ Hz, the
GW dominates the evolution of the BBH and the
eccentricity of the BBH reduces to $e_1\sim0.1$.
Alternatively, as the bottom panels indicate,
$f_{\rm GW}$ can be larger than $1$ Hz before
the GW dominates, and the eccentricity of the BBH is extremely high, i.e.,
$1-e_1\sim 10^{-3}-10^{-4}$.
We find that the latter case is very rare, i.e.,
$<0.1\%$ of all the BBH mergers in our simulation. However,
if the entering frequency can be as low
as about $\sim0.1$ Hz, then the probability can be much higher;
the inspiraling phase of BBHs, before GW dominating the evolution,
can now be observable (see top right panel of Figure~\ref{fig:klgw}).
If such phenomena can be detected, probably in decihertz GW detectors~\citep{decihertz18}, it can be a
strong evidence that those GW mergers are due to KL oscillations.

\subsection{Binary-single encounters}
\label{subsec:by_sn}

\begin{figure*}
\center
\includegraphics[scale=0.8]{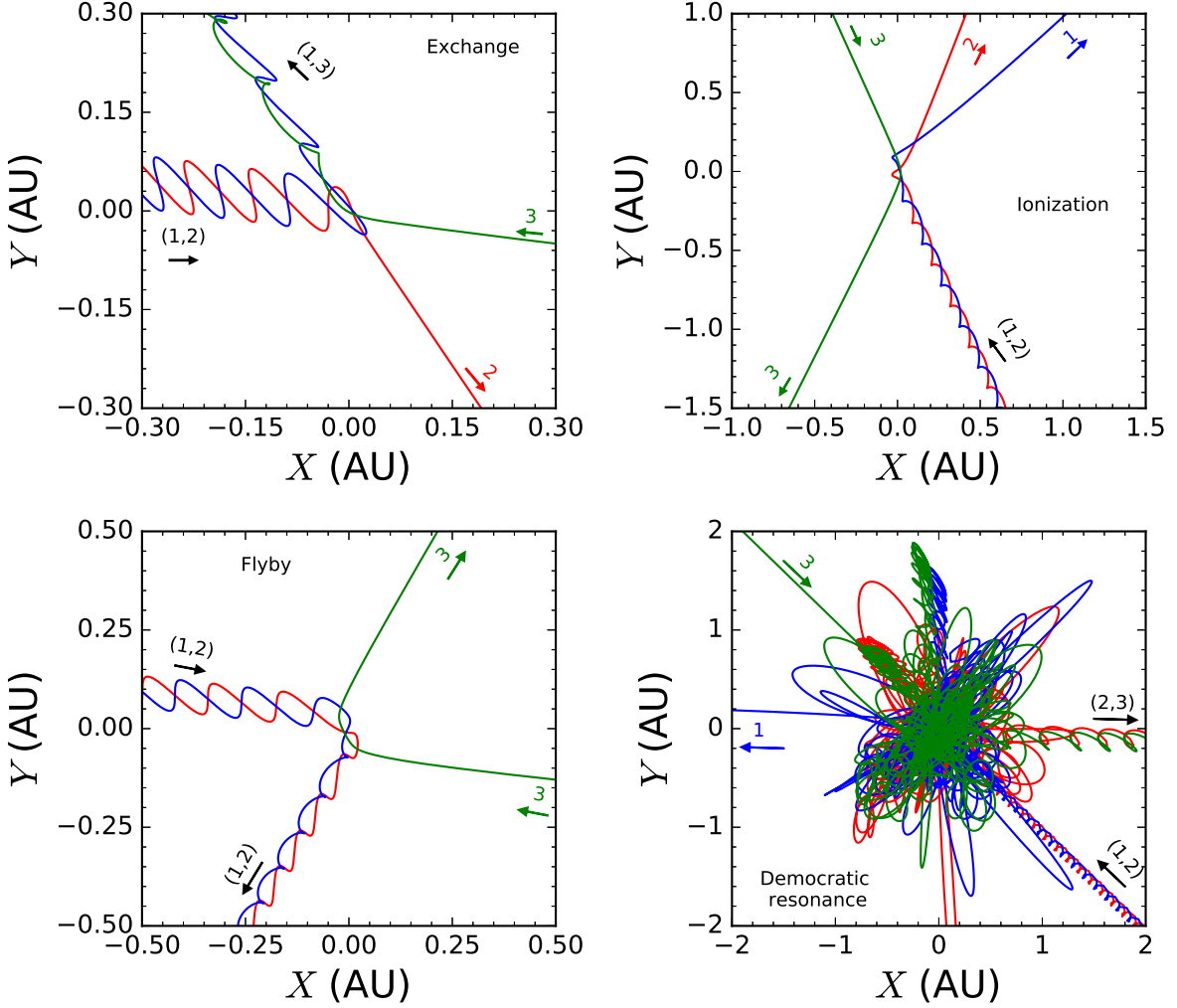}
    \caption{Four possible outcomes of an
    encounter between a BBH and a background star.  All panels show the
    trajectory of the BBH and the incoming star in the projected $X$ and $Y$
    axis.  The number $1$ and $2$ refer to the two components of the
    binary and $3$ refer to the incoming field star. In all panels the initial
  SMA of the BBH is $a_1=0.1 \, \AU$.}
\label{fig:track}
\end{figure*}

The BBH will experience encounters with the field stars (or stellar remnants)
if the stellar density is high. Such encounters can change both $a_1$ and
$e_1$, or even lead to exchange of the binary components.  There is a critical
velocity that defines the hard and soft regions of the binary
encounter~\citep{Hut83},
\be
v_c=\sqrt{\frac{Gm_T}{a_1}\frac{\mu_{12}}{m_3}},
\ee
where $\mu_{12}=m_Am_B/(m_A+m_B)$ is the reduced mass of the binary,
$m_T=m_A+m_B+m_3$ is the total mass, and $m_3$ is the mass of the
incoming star. Suppose that the velocity of the incoming star
with respect to the mass center of BBHs is $v_\infty$, if
$v_\infty>v_c$, the binary is soft, and is hard if $v_{\infty}<v_c$. After the
encounter, a hard binary always becomes harder, while
for a soft binary, it could become harder or softer, or even ionized
(disrupted).

The event rates of binary-single encounter depend on the density profiles,
whether the binary is hard or soft, and other details of the cluster. The
detailed calculations can be found in Appendix~\ref{apx:encounter}.
Assuming that $\alpha_\star=1.75$, for a soft binary, the event rate is given
by
\be\ba
R_{\rm EC}&=2.23\times10^{-7} ~{\rm yr^{-1}}~\Theta
\left(\frac{p}{0.1~{\rm AU}}\right)^2
\left(\frac{a_2}{10^3~{\rm AU}}\right)^{-9/4}\\
&\times
\left(\frac{\bh}{4\times10^6\msun}\right)^{7/8}\left(\frac{10~\msun}{m_\star}\right),
\label{eq:rec1}
\ea\ee
and for a hard binary, we have
\be\ba
R_{\rm EC}&=1.88\times10^{-9} ~{\rm yr^{-1}}~\Phi
\left(\frac{p}{0.1~{\rm AU}}\right)
\left(\frac{a_2}{10^4~{\rm AU}}\right)^{-5/4}\\
&\times\left(\frac{\bh}{4\times10^6\msun}\right)^{-1/8}
\left(\frac{10~\msun}{m_\star}\right)\left(\frac{m_T}{30~\msun}\right),
\label{eq:rec2}
\ea\ee
where $\Theta$ and $\Phi$ are dimensionless constants of
order unity, which depends slightly on the orbital eccentricity $e_2$ (See
Appendix~\ref{apx:encounter} for more details).  $p$ is the pericenter distance
of the incoming star with respect to the BBH, and it
relates to the impact parameter $b$ by
\be
b=p\sqrt{1+2m_TG/(v_\infty^2 p)}.
\label{eq:bp}
\ee
We can see that, for a typical galactic nucleus, the binary-single
encounter should be considered in the simulation, as they could be frequent,
especially in the inner regions of the nucleus cluster
(see also Figure~\ref{fig:ts}).

As the outcome of a binary-single encounter is complex and we find that
the change of the inner orbit of the BBH may have significant impacts
on their GW merging rates, such a process should be
calculated as accurately as possible.  Here we use explicit
three-body numerical integration to calculate the outcome of the binary-single
encounters. The incoming star is assumed coming from infinity with
$v_\infty=\sqrt{G\bh/(2a_2)}$ and impact parameter $0<b<b_{\rm max}$
which follows a distribution $f(b)\propto b$. Here $b_{\rm max}$
relates to $p_{\rm max}$ according to Equation~\ref{eq:bp}. We set
$p_{\rm
max}={\rm max}[6(m_\star/m_{\rm BBH})^{1/2},2]a_1(1+e_1/2)$
such that the simulation results can be converged.
As $m_\star/m_{\rm BBH}\sim 1$--$0.1$, $p_{\rm max}\sim 2\,a_1$--$9\,a_1$.  The initial
orbital orientation of the incoming star is set to be random. Initially the
incoming star is put at distance of $50\,a_1$ away from the BBH. After each
encounter, we determine the outcomes according to the relative energies and
separations between each particle (black holes or incoming stars).

Figure~\ref{fig:track} illustrates the four
most likely outcomes of binary-single encounters in our simulation: (1)
Exchange of one of the binary components with the incoming object. This event is
common if $m_{\rm BBH}\la m_\star$, i.e., when the incoming object is massive. The
incoming object is most likely a star considering that the number fraction of
black holes is very small ($\la10^{-2}$, see~\citet{Hopman06}), thus, the
exchange event usually leads to a non-BBH object.  In the current work,
we simply remove such binaries in the simulation. However, we notice that in
the future our work can be easily expanded to further consider the evolution of
 binaries of which one component being a star or a neutron star and the other
 being a black hole.
(2) Ionization, which means that the binary is destroyed due to the encounter.  This event is only
possible if $v_\infty>v_c$, i.e., when the binary is soft.  (3) Flyby, the
binary remains integrity after the encounter. However, the energy and
angular momentum of the inner binary, or equivalently, $a_1$ and
$e_1$, are both changed.  (4) Three bodies experience
chaotic evolutions, and finally, a binary is formed and the third star being
ejected from the system.  These events are only possible if $v_3<v_c$, i.e.,
when the binary is hard.  There are other possible outcomes, e.g., hierarchical
resonants~\citep{Heggie93}, however, only for hard binaries.  As most of the
binaries in our work are soft binaries, we do not discuss more details of them
here.

\subsection{The Monte-Carlo Schemes}
\label{subsec:MC_schemes}

We use the Monte-Carlo Schemes that are similar to~\citet{SM78} to
simulate the evolution of the inner and outer orbits of a BBH around
the MBH. Our method has combined all of the effects described
above, i.e., the NR and RR of the BBHs (Section~\ref{subsec:dy_outer}), the
tidal disruption of the BBHs (Section~\ref{subsec:tidal_disruption}), the KL oscillations and
the GW radiation of the inner orbit of the BBHs (Section~\ref{subsec:kl_gw}),
and the
binary-single encounters (Section~\ref{subsec:by_sn}).
The details of the method are described as follows.

Define $x=E/E_0$ as the dimensionless energy, where $E_0=-G\bh/r_{\rm h}$ is
the characteristic energy. We assume that
the energy distribution of the field stars is given by~\citep{LS77,
Magorrian99} $f(E)=(2\pi\sigma^2_0)^{-3/2}n_0 g_\star(x)$, where
\be
g_\star(x)=\frac{\Gamma(\alpha_\star+1)}{\Gamma(\alpha_\star-1/2)}
x^{\alpha_\star-3/2},~~3>\alpha_\star>1/2,
\label{eq:fe}
\ee
and $x>0$, where $\alpha_\star$ is the power law index of density profile of the field stars.
Here
$\sigma^2_h= G \bh/r_h$, $n_0$ is the number density of the stars at
position $r=r_h$. If $x<0$, we simply set $g_\star(x)=e^x$~\citep{LS77}.

We initially put all the BBHs near the edge of the influence radius of the MBH,
i.e., $r=r_{\rm i}=r_{\rm h}$ (or $x=0.5$).  The initialization of
the inner orbits of the binaries depends on the problems concerned, and
will be introduced in more detail at the beginning of
Section~\ref{sec:simulation} and Section~\ref{sec:event_rate}.
The dimensionless distribution function of the
energy of the BBHs' outer orbit is defined by:
\be
g_{\rm BBH}(x)=\frac{n(x)}{n(1)}\frac{\Gamma(\alpha_\star+1)}{\Gamma(\alpha_\star-1/2)}
x^{5/2},
\label{eq:gbbh}
\ee
where $n(x)$ is the number distribution of BBHs.
This is obtained according to $n(E)\propto f(E)E^{-5/2}$ in an isotropic cluster, where
$n(E)$ is the number density of BBHs as a funtion of the outer orbits' energy~\citep{Hopman09}.
In the above equation, $g_{\rm BBH}$ is normalized for the convenience of model comparison,
as $g_{\rm BBH}(1)=g_\star(1)$ for all models with the same $\alpha_\star$.

If $t$ is the current time of the simulation, we take the
following steps for each BBH:

\begin{enumerate}

\item A time step size $\delta t$ is determined according to
  Section~\ref{subsec:tstep}.

\item We trace the NR of the energy ($E=-G\bh/(2a_2)$) and NR and RR of the
  angular momentum ($J=\sqrt{G\bh  a_2(1-e_2^2)}$) of the outer binary by the
  method described in Section~\ref{subsec:dy_outer}.  The change of energy and
  angular momentum is given by
\be\ba
dE=&D_E^{NR} \delta t + y_1 \sqrt{D_{EE}^{NR} \delta t},\\
dJ=&\left(D_J^{NR} +D_J^{RR} \right)\delta t ,\\
&+y_2 \sqrt{D_{JJ}^{NR} \delta t}+y_3\sqrt{D_{JJ}^{RR}\delta t}.\\
\ea\ee
Here $y_1$, $y_2$ and $y_3$ are unity normal random variables. Note that $y_1$
and $y_2$ have correlations $\rho=D_{EJ}/\sqrt{|D_{EE}D_{JJ}|}$.  They can be
obtained by generating a bi-normal distribution and then taking
the transformation, $y_2\rightarrow y_1\rho+y_2\sqrt{1-\rho^2}$.
\item If the pericenter of the outer orbit of a BBH is within $r_p<3r_t$, and
  when ${\rm int}(t/P)-{\rm int}[(t-\delta t)/P]>0$, i.e., this BBH is near
  loss cone for an orbital period, we start a three
  body simulation, that traces the interactions between this BBH and the MBH
  (see Section~\ref{subsec:tidal_disruption} for more details).
\item The expected number of collisions between each BBH with the field stars is
  calculated according to $k=R_{\rm EC}\delta t$, where $R_{\rm EC}$ is given
  by Equation~\ref{eq:rbi}. The realized number of collisions, $n_{\rm EC}$, for each BBH
  is generated according to the Poisson distribution $P_{\rm EC}(k)$ (given by
  Equation~\ref{eq:pec}). We then take a number of $n_{\rm EC}$ successive
  three body integration to consider these encounter events
  (the details can be found in Section~\ref{subsec:by_sn}).
  Usually, the timestep is set such that $n_{\rm EC}\sim 1$ (see
  Section~\ref{subsec:tstep}).  Note that any change of the inner orbit of the
  BBH is saved for next steps.  After each encounter, the
  BBH is considered to be ionized if $\epsilon<10^{-3} m_\star
  |E|$, here $\epsilon=m_{\rm BBH}^2/(2a_1)$.
\item If the pericenter of the outer orbit of a BBH is $3r_t<r_p<{\rm
  min}[20, 0.58(T_{\rm GR}/P_2)^{2/3}]r_t$, we consider the KL oscillation of
  the inner binary that lasts for a time period of $\delta t$ (see
  Section~\ref{subsec:kl_gw} for more details).  We calculate the GW radiation
  timescale $T_{\rm GW}$ for this BBH at any moment, and the BBH is
  considered to be merged due to GW radiation if $\delta t-t'<50\,T_{\rm GW}$,
  where $0<t'<\delta t$ is the current time of simulation in the
  KL subroutines. We calculate the eccentricity and SMA of the BBH when
  it reaches $f_{\rm GW}=10$\,Hz according to Appendix~\ref{apx:GWfreq}.
  If $|\dot e_1^{\rm KL}|>|\dot e_1^{\rm GW}|$ and $f_{\rm GW}\ge10$\,Hz,
  the BBH is now entering the LIGO frequency band, and we record the
  current value of eccentricity and SMA of the BBH.

\item If $r_p>{\rm min}[20, 0.58 \, (T_{\rm GR}/P_2)^{2/3}]r_t$, we calculate
  the GW radiation of the inner binary according to Equation~\ref{eq:dae1_gw}.
  Similarly, we consider the BBH is merged if $10\,T_{\rm GW}<\delta t$, and
  we calculate the eccentricity and SMA of the BBH when it reaches
  $f_{\rm GW}=10$\,Hz according to Appendix~\ref{apx:GWfreq}.
\item We remove any BBH that has a dimensionless energy of $x<0.1$ or $x>10^4$.
  The inner boundary corresponds to a distance of about $10$\,AU if $\bh=10^6-10^7\msun$.
\item If the BBH is not destroyed (due to either merger of GW radiation or
  ionization, or tidal disruption, etc), repeat step 1
  to step 7 till the density profile of BBHs, i.e., $g_{\rm BBH}$,
  reach an equilibrium state\footnote{The equilibrium state is considered when the total
  time of simulation $t>T_{\rm NR}(r)$, where $T_{\rm NR}$ is given by
  Equation~\ref{eq:tnr}. Here $r=r_h/2$ if $\alpha_\star=7/4$ and $r=0.001r_h$ if
  $\alpha_\star=1$. In each simulation we output the density profile $g_{\rm BBH}(x)$
  at different time snapshot which usually is separated by about $\sim 0.1-0.5T_{\rm NR}$.
  The simulation continues until the $g_{\rm BBH}(x)$ of the last three snapshots converge.
  The last snapshot is considered as the equilibrium state and used for statistics.}.
  For any BBH that is destroyed, its simulation ends and the information is saved for statistics.
  The information include the inner and outer orbits of the BBH, the current simulation time, the
  number of different dynamical events it has been experienced, and etc.
\end{enumerate}

We trace the evolution of the ascending node
($\Omega_1$, $\Omega_2$), inclination ($I_1$, $I_2$) and the argument of periapsis
($g_1$, $g_2$) for both the inner and outer orbits of the BBHs.
Note that all these elements are defined with respect to an arbitrary selected
reference plane and a direction within it, which are the same for all BBHs in
a simulation. We assume that the vector RR process affects only the orientation of the
outer orbit ($\Omega_2$ and $I_2$), but not the inner orbit.
During the binary-single encounter, the orbital elements of the inner orbit are changed and
recorded for successive encounters. Currently, we do not consider the change of the outer orbit
due to the binary-single encounter, but defer it to future studies. Such simplification should not
significantly affect our results as the velocity dispersion in the cluster is usually much larger than
the velocity of the inner orbits of BBHs. All angles are considered and traced in the KL oscillations
and the BBH-MBH encounters. Note that the additional precession of $g_2$ due to distributed mass has
not been included in the current simulation (See Section~\ref{sec:discussion} for how it affects our results).

As sometimes the density of the binaries drops rapidly near the MBH, to
increase the number statistics we use the ``clone scheme'' similar
to~\citet{SM78}.
We generate a number of $\Pi-1$ clones of a BBH
(either an original or a clone one) when its energy
crosses the boundary $x=10^i$ from lower $x$, where $i=1,2,3$.
Here the number $\Pi$ is the amplification factor in the scheme, and is usually
a number selected between $2-10$ in each run, such that the number of BBHs in both outer
and inner regions of the cluster is sufficiently large for statistics.
If a clone particle crosses such boundary from higher $x$, it will be removed from the simulation (its information is not saved for statistics).
When a BBH is destroyed because of any step in (3)--(7), if it is the
original particle, a new particle (regarded also as a new original particle) at
$r=r_{\rm i}$ is generated. But if it's a clone particle, it is removed from
the simulation. Note that any BBH located at position $10^i<x<10^{i+1}$
has a statistical weight of $\Pi^{-i}$.

We use the code DORPI5 based on the explicit fifth (fourth)-order
Runge Kutta method \citep{DP80,Hairer93} to calculate the $3$-body dynamics
of the binary-single encounters and binary-MBH encounters, and the equations of the
motion of KL oscillation. The level of integration accuracy is always below
$10^{-12}$, which should be small enough for the convergence of the simulation results.
\subsection{Timesteps}
\label{subsec:tstep}

The timestep in our simulation should be small enough, such that the results
can converge. We use the step control similar to~\citet{SM78}.  The
timestep $\delta t$ satisfies
\be\ba
{\rm min}\left(\delta t\left|D_{E}^{NR}\right|,\sqrt{\delta
tD_{EE}^{NR}}\right)&\le0.15|E|,\\
{\rm max}\left(\sqrt{\delta tD_{JJ}^{NR}},\sqrt{\delta tD_{JJ}^{RR}}\right)
&\le{\rm min}[0.1J_{\rm c}, 0.4\left(1.0075J_c-J\right)],\\
{\rm max}\left(\sqrt{\delta tD_{JJ}^{NR}},\sqrt{\delta tD_{JJ}^{RR}}\right)
&\le{\rm max}\left(0.25|J-J_{\rm lc}|,0.1J_{\rm lc}\right).\\
\ea\ee
Also, if $r_p=a_2(1-e_2)<3r_t$, we set $\delta t\le P$, where $P$ is the
orbital period, such that the three body encounter between the BBH and the MBH
is considered for each period.  Under these conditions, the step size is small
enough, that the change of angular momentums is always smaller than the size of
the lose cone.

Also, if the binary-single encounters are considered, we require that the time step
size is small enough, that in each step the number of binary-single collisions
is of order unity,
\be
\delta t \le R_{\rm CO}^{-1}.
\ee
Here $R_{\rm CO}$ is given by Equation~\ref{eq:rbi}.  If the vector
relaxation is considered, to avoid a large change of the
orientation of the angular momentum, we set the time step such that
the change of angles is less than $30^\circ$, i.e.,
\be
\delta t \le 0.25 T_{\rm RR}^v.
\ee
where $T_{\rm RR}^v$ is given by Equation~\ref{eq:trrv}.

\section{Simulations: The effects of different dynamical processes}
\label{sec:simulation}

\begin{figure*}
\center
\includegraphics[scale=0.85]{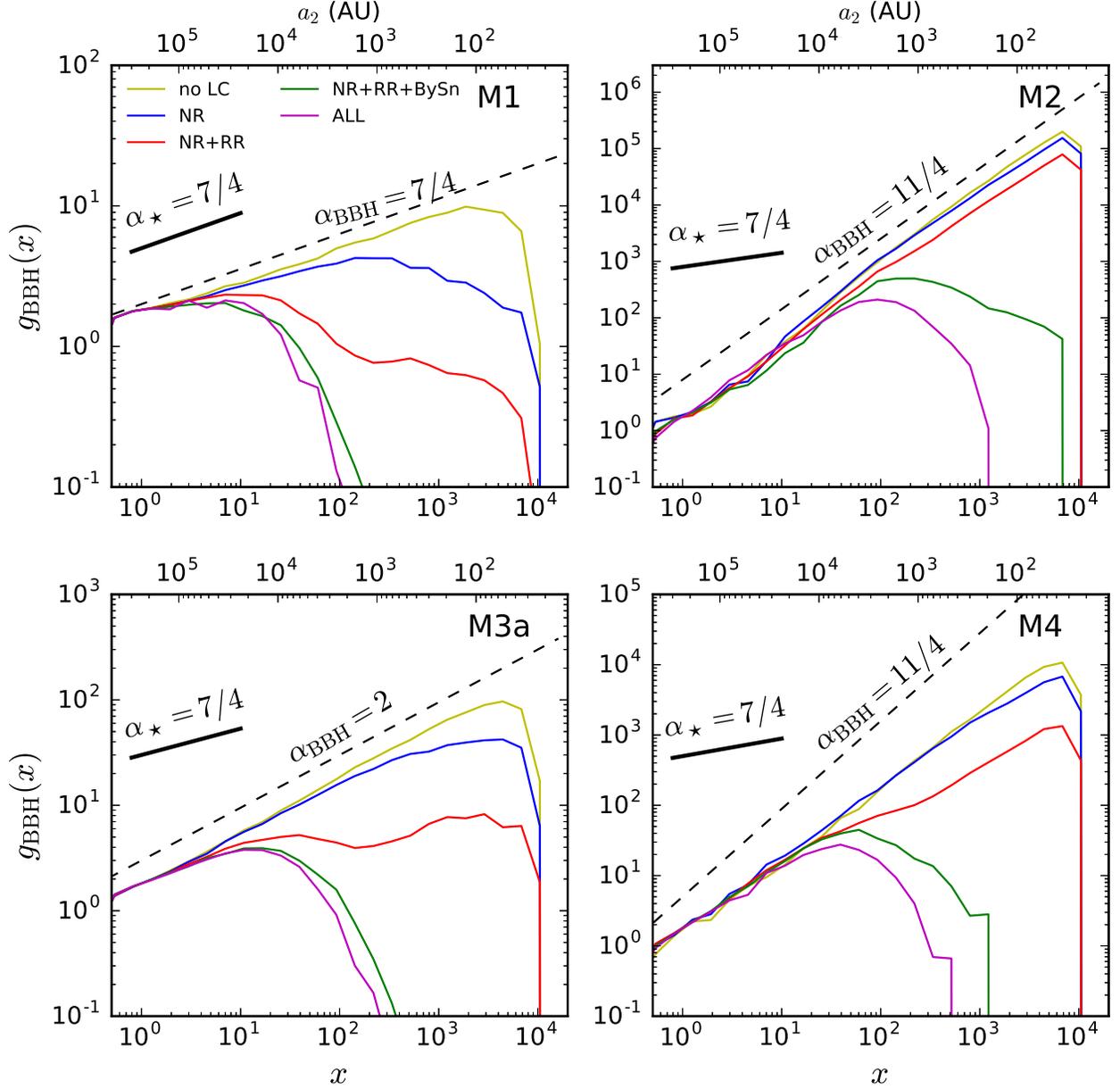}
\caption{The dimensionless distribution
function of the energy of the BBHs' outer orbit, i.e., $g_{\rm BBH}(x)$
(see Equation~\ref{eq:gbbh}) in four of the models in Table~\ref{tab:models}.
$x=E/E_0$ is the dimensionless energy. The lines in different colors
and linestyles show results when different dynamical processes
are being considered: The yellow solid line is the result
considering only the non-resonant (NR) relaxations and without loss cone; The
blue, red, green and magenta solid lines are the result considering the
loss cone effect.  The text in the legend show the effects being considered:
``NR'' means non-resonant relaxation, ``RR'' means resonant relaxation,
``BySn'' means the binary-single encounter, ``ALL'' means all of the above
effects, and additionally the KL effect, the GW orbital decay of the inner
orbit of BBHs. The black dashed line shows the
theoretical expectation for a no lose cone case, given by
Equation~\ref{eq:alpha_expect}, with the powerlaw index given in each panel.
The black thick solid line shows the density
profile of the background stars ($\alpha_\star=7/4$).}
\label{fig:ge}
\end{figure*}

\begin{table}
\caption{Models}
\centering
\begin{tabular}{lccccccccccc}\hline
& $m_{\rm BBH}$ ($\msun$) & $m_\star$ ($\msun$) & $a_1$ (AU) & $e_1$
& $\alpha_\star$ & $\xi^{a}$\\
\hline
M1                &  5$\times2$ & 10  & $0.1$  & $0 $   & 1.75  & $ 1.5$  \\
M2                & 10$\times2$ & 1   & $0.1$  & $0 $   & 1.75  & $ 59$  \\
M3a               & 10$\times2$ & 10  & $0.1$  & $0 $   & 1.75  & $ 5.9$  \\
M3b               & 10$\times2$ & 10  & $0.1$  & $0 $   & 1     & $ 7.5$  \\
M3c               & 10$\times2$ & 10  & $0.1$  & THM$^{b}$    & 1.75  & $ 5.9$  \\
M4                & 30$\times2$ & 10  & $0.2$  & $0 $   & 1.75  & $ 53$  \\
\hline
%
\end{tabular}
\tablecomments{
All the models have $\bh=4\times10^6\msun$.\\
$^{a}$ The initial value of $\xi$ for the binary at $r=r_h$ is given by
Equation~\ref{eq:xi}.\\
$^{b}$ Thermal distribution, i.e., $f(e) {\rm d}e=2e{\rm d}e$.
}
\label{tab:models}
\end{table}

In order to explore the effect on the density distribution of the BBHs and the
GW merge rates by different dynamical effects, we perform simulations according
to Section~\ref{sec:method} for some simplified models. In this section we fix
the MBH mass to be $4\times10^6\msun$ (which is the case of the Milky Way)
and fix the initial SMA of the inner orbit and mass of the
binary. For a more general and realistic initial condition that provides more
accurate estimations of the total GW event rates of BBH mergers for
all of the local galaxies, see Section~\ref{sec:event_rate}.

The initial conditions of some typical models are given in
Table~\ref{tab:models}. We assume two cases of the density profile of the field stars,
i.e., $\alpha_\star=7/4$, which corresponds to the value expected around a
MBH~\citep{BW76}, or $\alpha_\star=1$, which corresponds to a core-like cluster with
a shallower density profile.  For the models in Table~\ref{tab:models},
initially the mass of the background particle is $m_\star=1\msun$ or $10\msun$.
The former one assumes that later-type stars dominate the central
regions while the latter one assumes that black holes
dominate. We assume that the two components of BBH are of equal mass, and with mass $5\msun$, $10\msun$ or $30\msun$ in different models. In
the case of $m_A=m_B=5\msun$ the mass of the BBH equals to the field
stars, thus the density profile of them should follows also the cusp profile,
i.e., $\alpha_\star=7/4$.  We consider the case of $m_A=m_B=30\msun$ such that
they are similar to the masses of the merged BBH that was observed by the
Advanced LIGO, e.g., GW150914~\citep{Abbott16a} and
GW170104~\citep{Abbott17a}. Initially the SMA of the BBH is fixed to
$a=0.1\AU$ or $a=0.2\AU$. The initial eccentricity follows a thermal
distribution $f(e_1)\propto e_1$, or we simply set $e_1=0$.  To reduce the
effect of GW orbital decay by the binary itself, we set the maximum of $e_1$
such that the binaries have a long GW decay time, i.e., with $T_{\rm GW}>1$
Gyr.

We assume that the initial eccentricity of the outer orbit of the BBHs
follows $f(e_2)=2e_2$, and a random initial orientation of both the inner and
the outer orbits of the BBHs.  For each model, we perform a series of
simulations according to the method described in Section~\ref{sec:method}, with
all or some of the dynamical effects being considered.  We can
turn on the dynamical effects that we are interested in, and then
off, and see the effect of the difference on the results. The results are
shown in the following sections.

\subsection{The dynamical evolutions of the BBHs}
\label{subsec:dynamics_BBHs}
\begin{figure*}
\center
\includegraphics[scale=0.8]{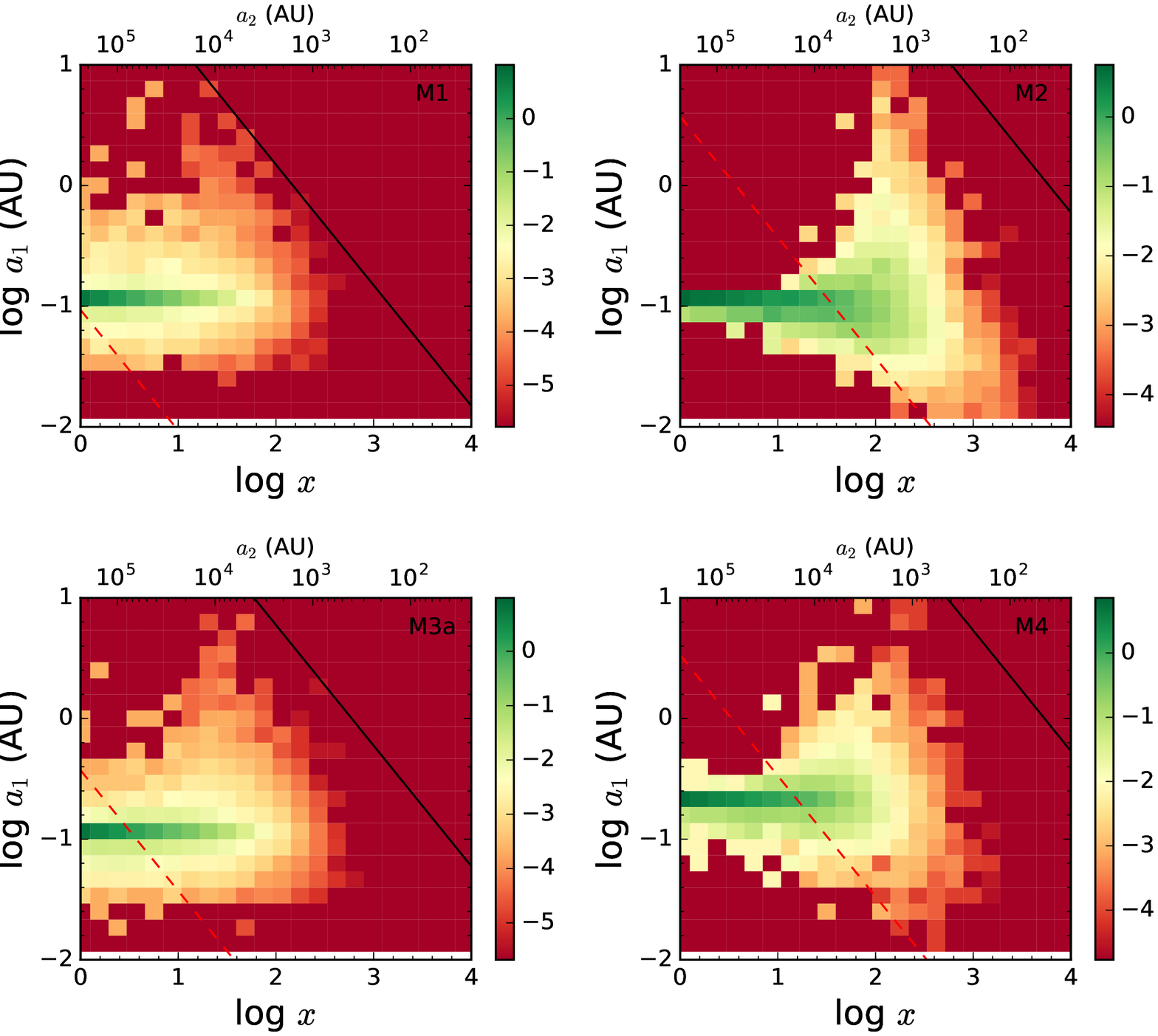}
\caption{The distribution of the BBHs in the $\log a_1$--$\log x$ space for
  models M1, M2, M3a and M4. Here $x=E/E_0$ is the dimensionless
  energy.  The color contours show the number density of BBHs in log
  scale per dex$^{-2}$. Note that the non-resonant, resonant relaxations and the
  binary-single encounters are considered, but other effects (KL
  and GW orbital decay) are ignored. The red dashed line
  shows the position where $\xi=1$, which separates the soft (above the
  line) and hard (below the line) binary regions.  The black solid line corresponds to
$\epsilon=10^{-3}m_\star |E|$. In regions above this line, the binary is considered to be too soft to
exist.}
\label{fig:ainE_ion}
\end{figure*}
\begin{figure*}
\center
\includegraphics[scale=0.8]{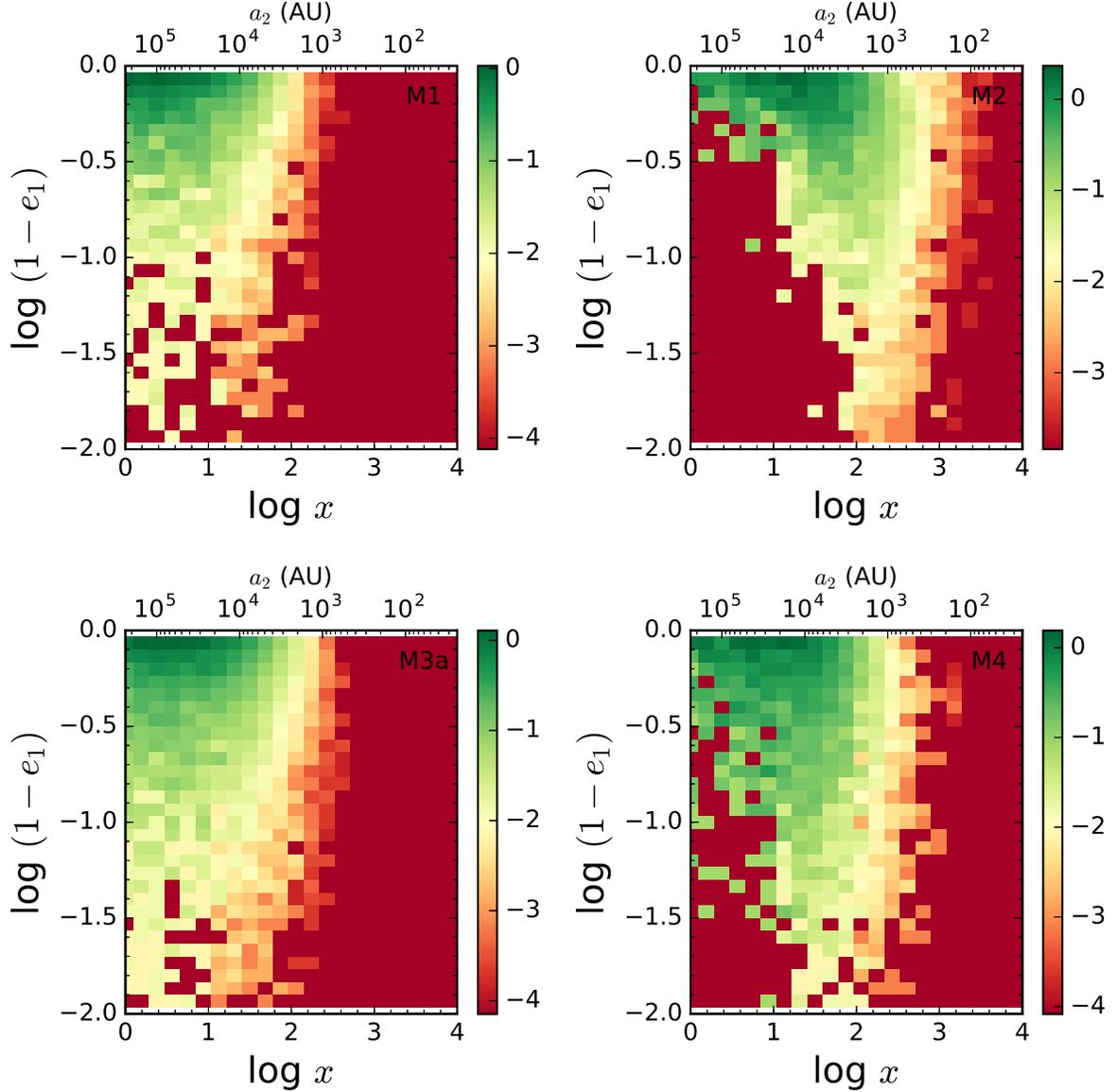}
\caption{Similar to Figure~\ref{fig:ainE_ion} but for $\log (1-e_1)$--$\log x$.
Here $e_1$ is the eccentricity of the inner orbit of the BBHs.}
\label{fig:einE_ion}
\end{figure*}

As the number density is dominated by the field stars, the expected density
profile of the BBHs should be the same as that of the field stars if
$m_{\rm BBH}=m_\star$ and show mass segregation effects if
$m_{\rm BBH}>m_\star$. The expected index of the
density profile is given by~\citep{Alexander09}
\be\ba
\alpha_{\rm BBH}&=\left\{\begin{array}{cc}
3/2+m_{\rm BBH}/(4m_\star)&{\rm if}~m_{\rm BBH}/m_\star\la4\\
9/2-\alpha_\star &{\rm if}~m_{\rm BBH}/m_\star\ga4
\end{array}\right.
\label{eq:alpha_expect}
\ea
\ee
The yellow solid lines in Figure~\ref{fig:ge} show the
distribution function $g_{\rm BBH}(x)$ in model M1, M2, M3a, and M4 when
only the NR is considered, and without the loss cone.  For model M1, in which
the mass of the binary $m_{\rm BBH}=10\msun$ is equal to the field stars, the density
profile of BBHs is the same as that of the field stars, i.e., $\alpha_\star=7/4$,
consistent with Euqation~\ref{eq:alpha_expect}. For models M2, M3a, and
M4, the density profile of the BBHs also follows
Equation~\ref{eq:alpha_expect}, consistent with the theoretical
expectations.

When additionally the loss cone is considered, the results of density profile in
different models are shown with the blue solid lines in
Figure~\ref{fig:ge}. We can see that the loss cone can
reduce the number of BBHs in the inner regions.  The RR process can cause
additional decrease of BBHs in the inner regions (see the red solid
lines in Figure~\ref{fig:ge}), as also suggested in other
studies~\citep[e.g.,][]{Hopman06}. This is because the RR can excite the
eccentricity of the outer orbit of BBHs and move them more efficiently to the
loss cone regions.  The decrease of BBHs due to the NR or RR
effects is only significant when $m_{\rm BBH}\simeq m_\star$.  If $m_{\rm BBH}\ga
4 m_\star$, the BBHs destroyed in loss cone are quickly replenished by the BBHs
moving from other regions under the strong mass segregation effect.

The binary-single encounters can reduce further the number of BBHs in the inner
region (see the green lines in
Figure~\ref{fig:ge}), mainly because the BBHs in the inner regions become soft
binaries in all models. We define a dimensionless parameter
$\xi$~\citep{Hopman09},
\be
\xi=\frac{\epsilon}{m_{\star}\sigma^2}\simeq
\frac{m_A m_B a_2}{2m_\star \bh a_1}.
\label{eq:xi}
\ee
Then the binary is soft if $\xi\ll1$ and hard if $\xi\gg1$.
Table~\ref{tab:models} shows the value of $\xi$ at
$a_2=r_h$.  We can see that for models M1 and M3a-c, within the cluster
the BBHs are mostly soft binaries.
For models M2 and M4, the binary is hard near the edge of the cluster.
However, they will be soft in the inner regions of the cluster.

The soft binaries are likely to be ionized after encounter.
Figure~\ref{fig:ainE_ion} shows the distribution of the SMA of the
inner orbit of the BBHs for different models (note that KL and GW
orbital decay are still ignored in these figures). We can see that in the hard
binary region, i.e., $\xi>1$, the binary always becomes harder after
encounter. However, in the soft binary region, the binary-single
encounters can increase the SMA ($a_1$) of the inner binary.  For model M2, as
$m_{\rm BBH}\gg m_\star$, a significant number of BBHs sink into the inner regions of
the cluster under strong mass segregation effect, where the rates of
binary-single encounters are dramatically increased. In all models, in the soft
binary region, such encounters can still decrease $a_1$ down to $10^{-2}$\,\AU,
which will help increase the event rates of GW mergers of the BBHs.

Similarly, Figure~\ref{fig:einE_ion} shows the distribution of the
eccentricity of the inner orbit of the BBHs. We can see that in all models, the
eccentricity of the binaries will be dramatically changed due to the
binary-single encounters. The maximum eccentricity can be up to $\ga 0.99$, which
will further increase the event rate of GW mergers of the BBHs.

The magenta lines in Figure~\ref{fig:ge} show the density distribution of BBHs
when all of the dynamical effects (including the vector RR relaxation, KL oscillations
and the orbital decays due to GW radiations) are considered. Comparing the
green lines with the magenta lines in Figure~\ref{fig:ge}, we find that
considering the GW orbital decay and KL effects can reduce further the number
of BBHs in the inner regions, which means that many BBHs are merged.
We find that the vector RR relaxation process has insignificant effects
on the number density of BBHs. In all models, the BBHs can survive  if
they are $10^2$--$10^4\AU$ away from the MBH Only in the model M2, where the mass
segregation effects are the most significant, the BBHs can sink into more inner region
with the distance of $\sim100\,\AU$ from the MBH.

We find that the density profile of BBHs in models with $\alpha_\star=1$
(e.g., M3b)
is similar to those of $\alpha_\star=7/4$, and only in the case that
$m_{\rm BBH}\gg m_\star$, we find that the BBHs are more concentrated in the central
regions for the case $\alpha_\star=1$. These simulation results are
consistent with what is expected in Equation~\ref{eq:alpha_expect}.

\subsection{GW mergers of BBHs}

\begin{figure*}
\center
\includegraphics[scale=0.85]{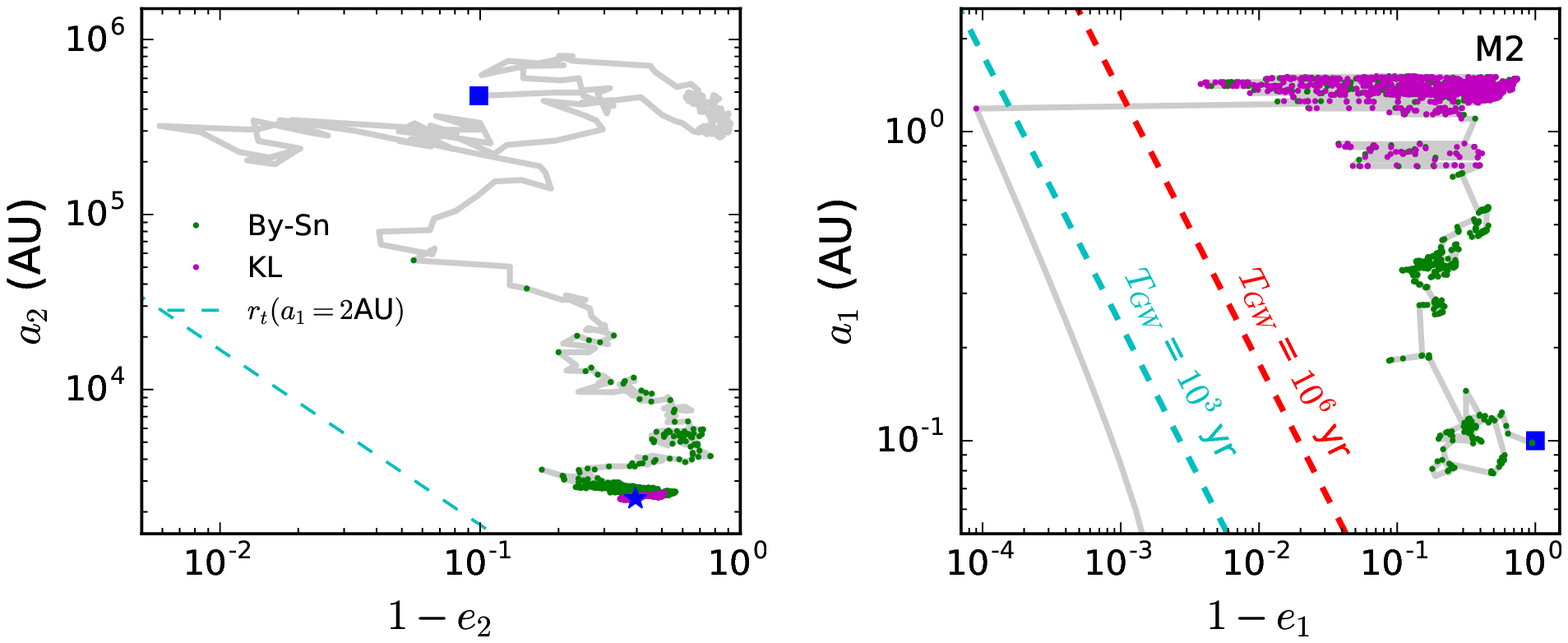}
\includegraphics[scale=0.85]{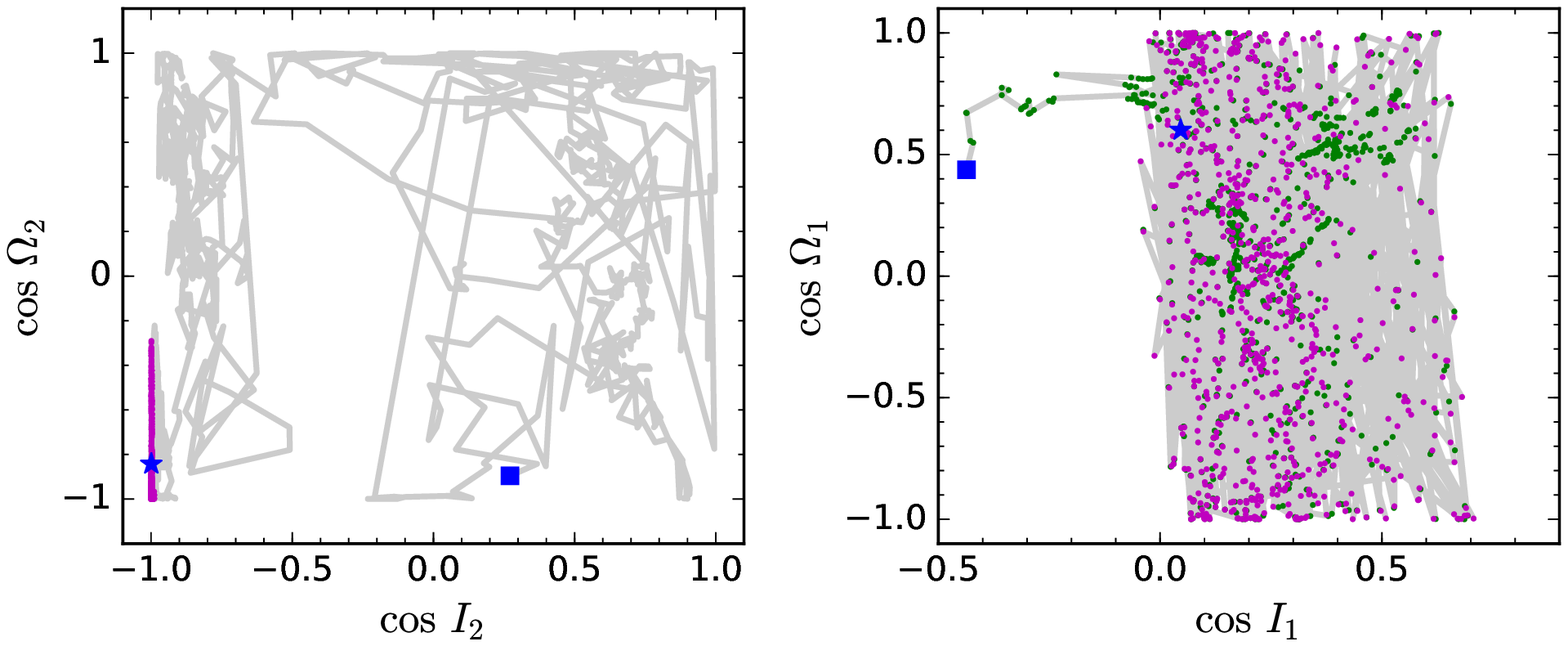}
\caption{ An example of evolutions of BBH's inner and outer orbits.
Left panels: The evolutions of the outer orbit of the BBH.
  The dashed cyran line in the top left panel shows the tidal
  radius for the BBH with $a_2=2\,\AU$.  In the bottom left panel, the
  evolution of $\Omega_2$ and $I_2$ are mainly driven by vector RR.
Right panels: The evolutions of the inner orbit of the BBH.  The dashe lines  in the top right panel
 show the GW timescales. ``By-Sn'' means the position after the binary-single encounter;
  ``KL'' means the position after the KL oscillations.
  The filled blue square symbol marks
  the starting position of the BBHs in each panel. The filled blue star symbol in all panels
  marks the position where the BBH is merged.
  The initial condition of the model is given in model M2.
}
\label{fig:evlM2}
\end{figure*}

\begin{figure*}
\center
\includegraphics[scale=0.85]{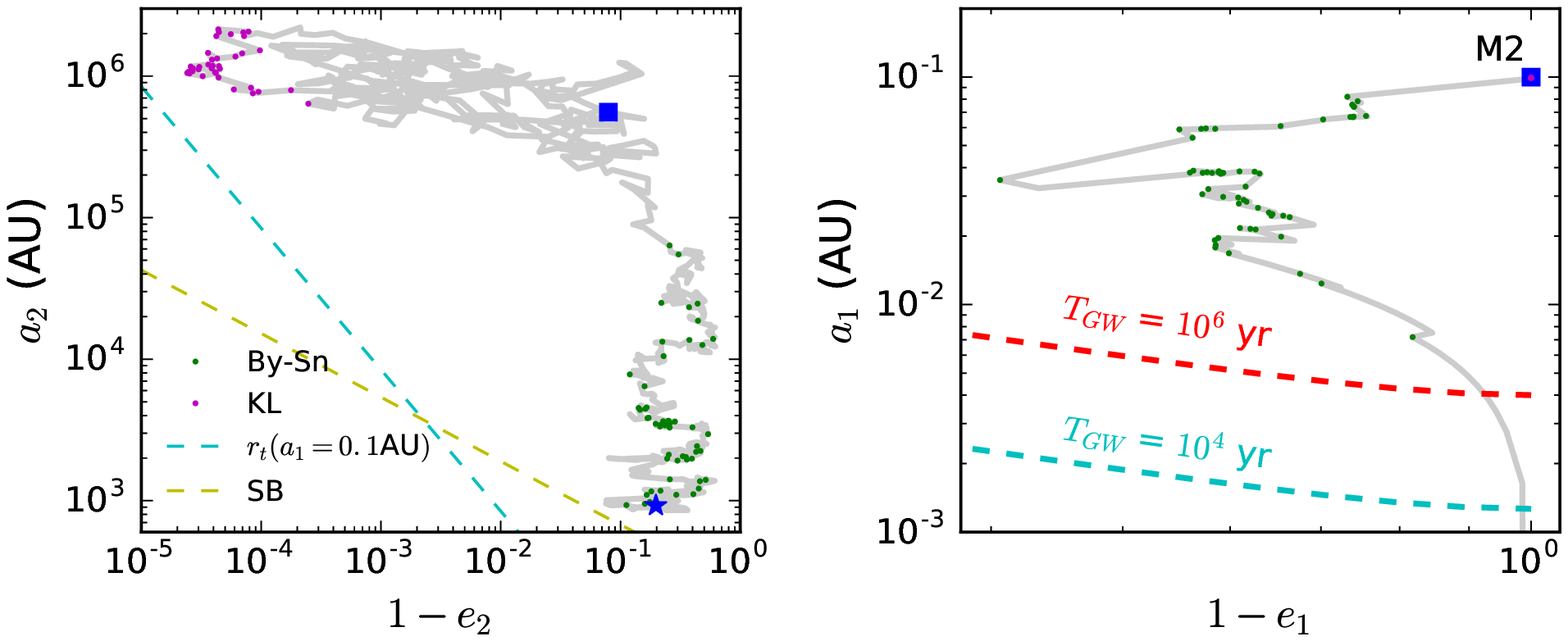}
\includegraphics[scale=0.85]{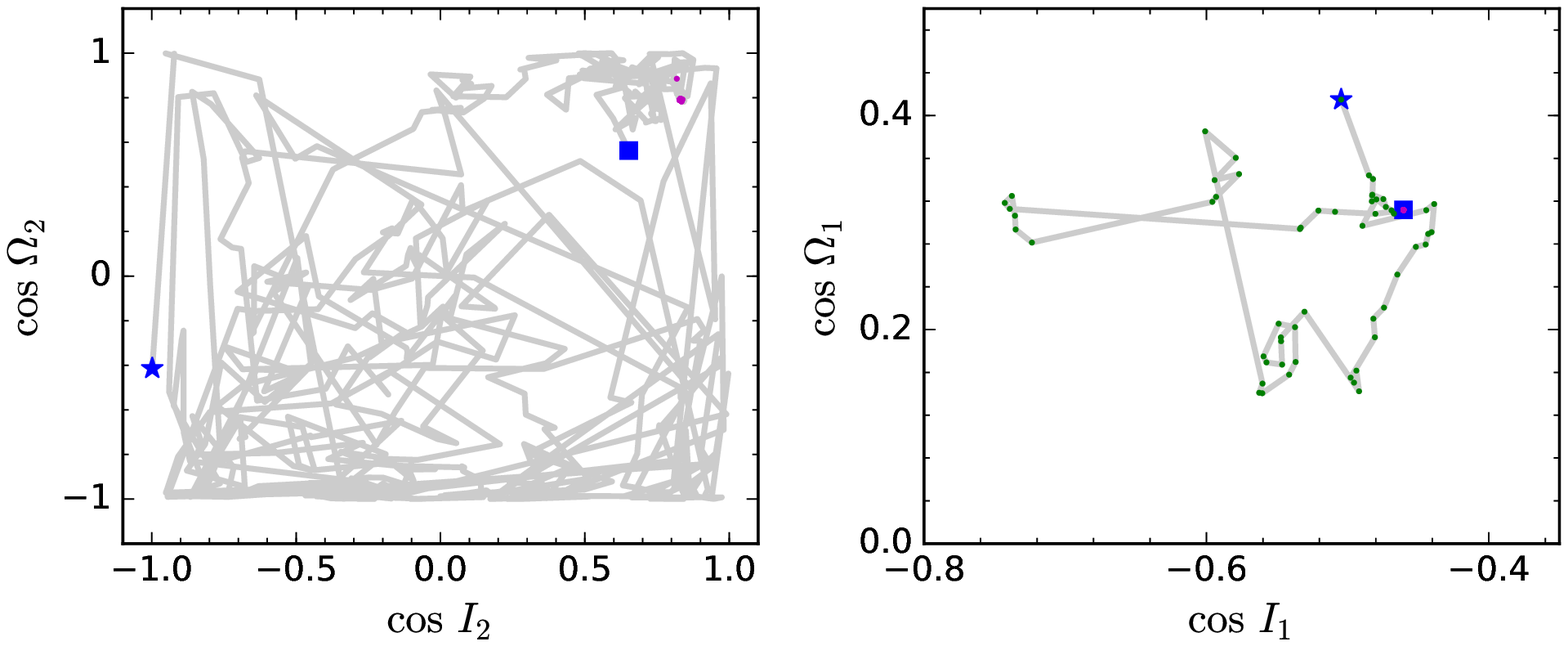}
\caption{Similar to Figure~\ref{fig:evlM2} but for a different BBH, where
the initial condition is given by model M2. The dashed yellow line
shows the Schwarzschild barrier, below which the RR is suppressed.}
\label{fig:evlM22}
\end{figure*}

\begin{figure*}
\center
\includegraphics[scale=0.85]{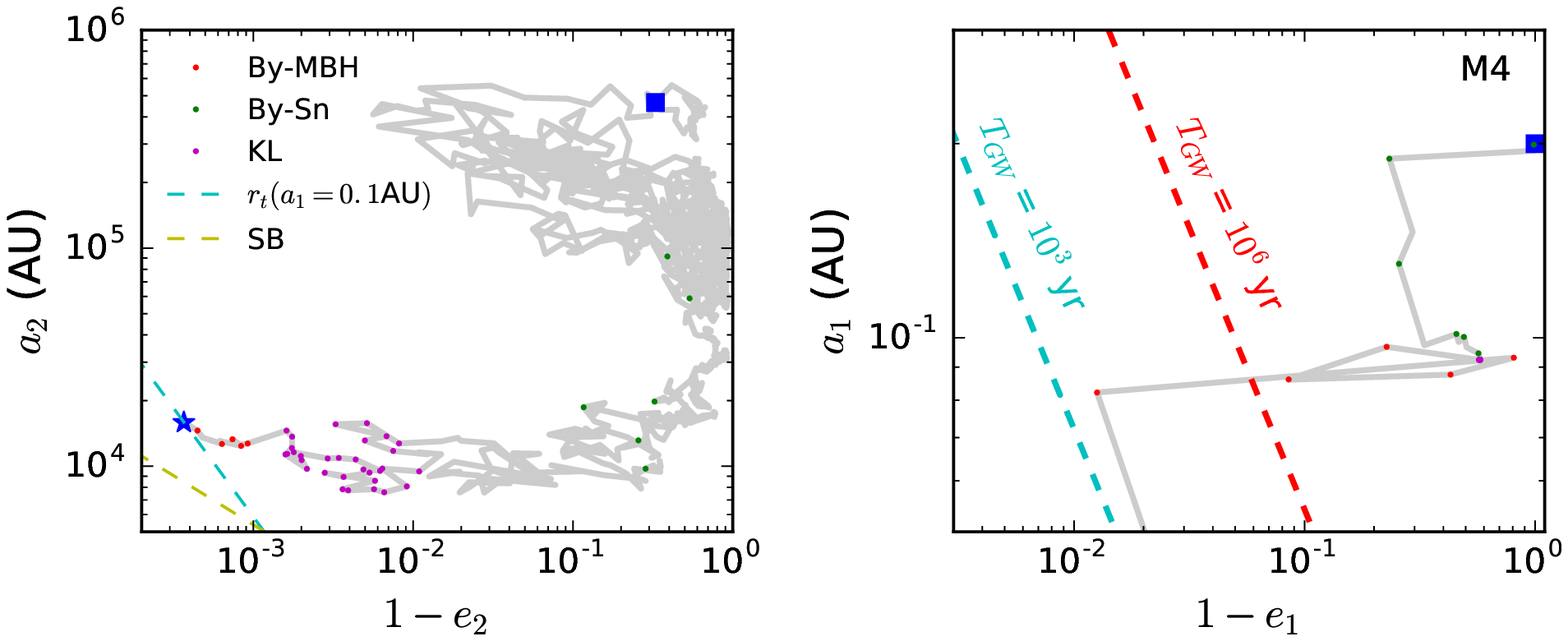}
\includegraphics[scale=0.85]{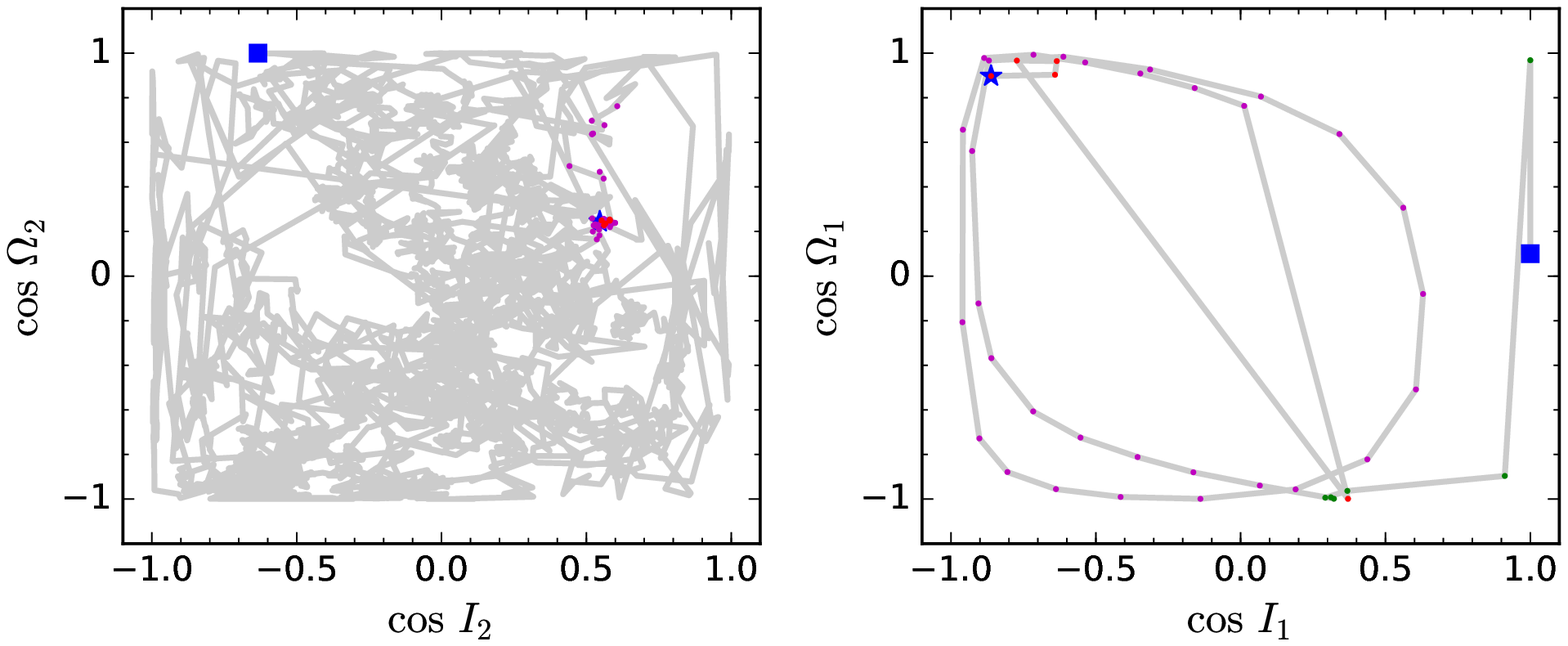}
\caption{Similar to Figure~\ref{fig:evlM2} but for a different BBH, where
the initial condition is given by model M4.  ``By-MBH'' means the position
after the binary-MBH encounter. The dashed yellow line
shows the Schwarzschild barrier, below which the RR is suppressed.}
\label{fig:evlM4}
\end{figure*}

There are a number of possible dynamical channels to merge BBHs in the cluster
containing a MBH. Here we exemplify three of them, which are presented in
Figures~\ref{fig:evlM2}--\ref{fig:evlM4}. The details are described as
follows.

In the first case, the BBHs experience hundreds to thousands times of
binary-single encounters, and then merge under the combined effect of KL
oscillations and the binary-single encounters. This channel is more common for
those models with strong mass segregation effects, i.e., $m_{\rm BBH}\gg m_\star$.
Figure~\ref{fig:evlM2} shows an example of the evolutions of the BBH's
inner and outer orbits in model M2.  The outer orbit of the BBH first gradually
migrates from the outer parts of the cluster ($a_2\sim10^5 \, \AU$) into the inner
regions ($a_2\sim10^3\,\AU$).  Along the way the binary-single encounters become
more and more frequent.  Each binary-single encounter changes the inner orbit
slightly, as the incoming field star is much lighter than the
components of the BBHs, i.e., $ m_A, m_B\gg m_\star$. From the top
right panel of Figure~\ref{fig:evlM2}, we can see that the encounters gradually
increase the $e_1$ of the BBH up to $\sim0.8$--$0.9$, and the SMA of
the BBH also increases up to about $2\,\AU$. When the pericenter of the BBH's
outer orbit approaches to about tens of $r_t$, the KL effects
become important, which can increase $e_1$ rapidly. Note that
simultaneously the binary-single encounters also change the inner
orbits. Under these two effects, the inner orbit of the BBH finally
reaches to a very high eccentricity and eventually the BBH merges due to GW
radiations.

The bottom panels of Figure~\ref{fig:evlM2} show the
evolutions of the orbital orientations of both the inner and outer orbits of the
BBH. The orientation of the outer orbit evolves under
vector RR and the KL effect (bottom left panel of Figure~\ref{fig:evlM2}).
The orientation of the inner orbit is changed cumulatively
during multiple binary-single encounters, and becomes quite rapidly during the KL
precesses (bottom right panel of Figure~\ref{fig:evlM2}).
The rapid oscillations of the inner orbital orientations can help to
trigger the merging of BBHs by KL effect. For example, before merging,
$I_2\sim \pi$ and $I_1\sim \pi/2$, the relative inclination between the
inner and outer orbits is close to $\pi/2$, which can help to enhance the eccentricity
oscillations and merging the BBHs~\citep{Wen03}.

Figure~\ref{fig:evlM22} shows another example of the BBH's evolutions
in model M2.  In Figure~\ref{fig:evlM22}, the SMA of the BBH always
tends to decrease due to multiple binary-single encounters. This is
because the BBH is mostly remaining in the hard binary regions,
and the binary-single encounters always decrease the SMA for a hard binary.  A
hard binary makes the frequency of binary-single encounters drops and prevents
another encounter even if it reaches the inner regions.

Different from model M2, M3a assumes more massive
background field stars. Such initial conditions have two consequences: (1) The
change of the inner orbit of the BBHs can be dramatic. (2) The number of
encounters is much smaller in M3a than in M2, as the number of
field stars is much smaller (due to their massiveness) according to
Equations~\ref{eq:rec1} or~\ref{eq:rec2}.  In these cases, the BBHs can
experience only a few binary-single encounters, which will dramatically
increase the eccentricity of the binary, and soon lead to merger
due to the GW radiations.

In some rare cases, the GW merging event can be triggered due to
the multiple encounters between the BBH and the MBH, as shown in
Figure~\ref{fig:evlM4}. In the final revolution,
from the bottom panels of Figure~\ref{fig:evlM4} we find that
the relative inclination between the plane of the inner and outer orbits is
$\sim\pi/2$. As the tidal force of MBH is nearly perpendicular to the orbital
plane of the BBH, it tends to suppress the BBH~\citep{Zhang10}
and help to merge the BBH.

This usually happens when the BBH is at the outer parts
of the cluster, such that it can approach to the loss cone region without being
ionized by the binary-single encounters, and that the effect of KL oscillations
is not so significant. However, we notice that if the
binary-single encounters are switched off, then this case is very frequent. The
eccentricities of the BBHs can be excited to very high values
due to the multiple encounters between BBH and MBH, and the GW is so
strong that they can merge before the next encounters with the MBH.

In all models, we find that the SB does not affect much the steady-state
flux of binaries into the loss cone, which may be important for those single stars~\citep{Baror16}.
As we can see from Figures~\ref{fig:evlM22} and~\ref{fig:evlM4},
in most cases SB is below the tidal radius of BBHs with $a_1\ga0.1\AU$,
and the SB is only effective if $a_2\la300\AU$, where most BBH can not penetrate into.
As RR is effective below $a_2\sim10^4\AU$ (See Figure~\ref{fig:ts}),
the tidal rates of BBHs are not affected much by the SB, and the RR process
can still drive most of the BBHs effectively into loss cone regions.

At the end of the simulation, suppose that the number of
integrated binaries that survived in the simulation is $N$, and the rate of merging events
is $\dot N_e=dN_e/dt$, where $dN_e$ is the total number
of merging events during time span $dt$. We can
define a normalized merging event rate in the
nucleus cluster, $\mathscr{R}=\dot N_e/N$. It means that if we
observed only one BBH in the real galactic nucleus
clusters at current moment, the merging event rate in
such cluster is then given by $\mathscr{R}$.  Note here that $N$ and $\dot N_e$
are calculated after considering the weight of each event in the clone
scheme (see Section~\ref{subsec:MC_schemes}).

Based on $\mathscr{R}$, the merging rate of the BBHs in a real galactic
nucleus can then be estimated after some scaling with the realistic number of BBHs
in each galactic nucleus. If assuming a constant number fraction, i.e., $f_{nb}$,
of BBHs in the cluster with respect to the field stars, the event rate is given by
\be
R=\frac{f_{nb}\bh}{m_\star}\mathscr{R}.
\label{eq:raten}
\ee
If alternatively assuming a constant mass fraction, i.e., $f_{mb}$, of BBHs
with respect to the field stars, it becomes
\be
R=\frac{f_{mb}\bh}{\langle m_{\rm BBH}\rangle}\mathscr{R},
\label{eq:ratem}
\ee
where $\langle m_{\rm BBH}\rangle $ is the mean mass of the BBHs.

In this work, we assume that the mass and the number fraction of
all stellar black holes are much smaller than the field stars. If assuming
a constant mass fraction, we set $f_{mb}=10^{-3}$. If assuming
a constant number fraction of BBHs, we set
$f_{nb}={\rm min}(10^{-2}m_\star/\langle m_{\rm BBH}\rangle,10^{-3})$, such that the
total mass of the BBHs does not exceed $\sim 0.01 \, \bh$, which is likely the mass
fraction of stellar-mass black holes in a nucleus cluster under standard
initial mass function of stars~\citep{Hopman06}.

\begin{table}
\caption{Dependence of the merging rate of BBHs on different dynamical effects}
\centering
\begin{tabular}{lccccccccccc}\hline
 model  & ALL        & Vec RR off   & KL off & By-Sn off \\
\hline
M1    & $3.3\pm0.2^a$   & $2.5\pm0.2$  & $2.6\pm0.2$  &  $1.5\pm0.2$   &    \\
& $3.3\pm0.2^b$   & $2.5\pm0.2$  & $2.6\pm0.2$  &  $1.5\pm0.2$   &    \\
M2 & $1046\pm76$      & $939\pm72$   & $899\pm70$
&  $362\pm48$  &  \\ &
$104.6\pm7.6$    & $93.9\pm7.2$ & $89.9\pm7.0$
&  $36.2\pm4.8$  &   \\
M3a   & $11.1\pm0.3$    & $11.0\pm0.3$ & $10.5\pm0.3$ &  $3.3\pm0.2$   &    \\ &
$5.5\pm0.1$    & $5.5\pm0.1$ & $5.3\pm0.1$   &  $1.7\pm0.1$   &    \\ M3b   &
$9.0\pm0.4$    & $7.8\pm0.4$   & $8.2\pm0.4$ &  $3.3\pm0.2$   &    \\
      & $4.5\pm0.2$     & $3.9\pm0.2$  & $4.1\pm0.2$  &  $1.7\pm0.1$   &    \\
M3c   & $22.1\pm0.9$    & $19.6\pm0.8$   & $20.3\pm0.8$ &  $6.9\pm0.5$   &    \\
      & $11.0\pm0.4$    & $9.8\pm0.4$   & $10.1\pm0.4$ &  $3.4\pm0.2$   &    \\
M4    & $169\pm10.6$      & $175\pm11$   & $163\pm10$   &  $48\pm5.7$  &   \\
      & $28.1\pm1.8$      & $29.3\pm1.8$   & $27.3\pm1.7$   &  $8.0\pm1.0$  &   \\
\hline
%
\end{tabular}
\tablecomments{
Note that the rates in this table are only for the purpose of demonstrating the
effects of different dynamical processes. For each model, we have assumed
Milky-Way-like galaxies with MBH $\bh=4\times10^6\msun$,
and have oversimplified initial conditions of the BBHs (see Table~\ref{tab:models}).
For more realistic event rate estimations in the local universe,
see Section~\ref{sec:event_rate} or
Table~\ref{tab:cmodels}.\\
  $^a$Merging rates in unit of Gyr$^{-1}$, assuming
  a constant number fraction of BBHs, i.e., $f_{nb}={\rm
  min}(10^{-3},10^{-2} m_\star/\langle m_{\rm BBH}\rangle)$\\
  $^b$Merging rates in unit of Gyr$^{-1}$, assuming
  a constant mass fraction of BBHs, i.e., $f_{mb}=10^{-3}$
}
\label{tab:GW_rate}
\end{table}

Table~\ref{tab:GW_rate} shows the estimated GW event rates according to
Equation~\ref{eq:raten} and Equation~\ref{eq:ratem} in different models. The first row for each
model shows the result if assuming a constant number fraction of BBHs and
$f_{nb}={\rm min}(10^{-3},10^{-2} m_\star/\langle m_{\rm BBH}\rangle)$ while
the second row assuming
a constant mass fraction of BBHs and $f_{mb}=10^{-3}$.  We can see that, if all the
dynamical effects are considered,  the event rates
range from $1$--$10^3$ Gyr$^{-1}$ in the six models investigated here. Merging
rates assuming a constant number fraction are
usually larger than those assuming a constant mass fraction, as usually
the field stars are considered lighter than the
BBHs.  The change of $\alpha_\star$ parameter does not affect significantly the
merging rates, as it changes only the density profiles
in the cases that $m_{\rm BBH}\gg m_\star$. If alternatively the initial eccentricity
is not zero, but follows the thermal distribution, the merging
rates will be about twice larger. This can be straight forwardly obtained as
the larger the eccentricity, the easier the BBHs can be merged.

We find that the exchange of the BBH with the incoming stars is common in some
models, e.g., M1, M3a, as usually the exchange event is frequent if
$m_\star\sim m_{\rm BBH}$.  These exchange events decrease the total GW event
rate significantly. By removing the exchanged BBHs from the simulation, we find
that the GW event rate is about twice smaller.

As we have shown in Figures~\ref{fig:evlM2}--\ref{fig:evlM4}, the
mergers of BBHs are results of the combinations of various dynamical
effects. To explore their importance in the merger of BBHs, we also
run similar simulations for each model, but excluding some dynamical
processes.  The results are shown in Table~\ref{tab:GW_rate}. We do
not see a clear difference when the vector RR effect is being considered or
not. If the vector RR effects are
switched off, the event rates usually drop only slightly for all
models.  As we consider only $\bh=4\times10^6\msun$ in this section, this is
consistent with~\citet{Hamers18}, that the vector RR effect is more important for MBHs
with $\bh=10^4$--$10^5\msun$, but quickly becomes negligible for
massive MBHs.

Many of the BBHs are ionized or tidally disrupted along the way from the outer to
the inner regions.  Thus, the KL effect does not contribute much to
the total merging rate as few of the BBHs can survive to the
vicinity of MBH where it is most efficient. If the KL effects
are switched off, we do not see a significant difference on the GW merging rates
(See Table~\ref{tab:GW_rate}).  On the other hand, many of the BBHs may
coalesce due to multiple binary-single encounters.  Our results suggest that
the binary-single encounters can be an important channel that leads to the
mergers of the BBHs.

If we switch off the binary-single effects, then the inner orbits of
BBHs are affected only by KL oscillations or the strong encounters with the MBH.
We find that in these cases, most of the BBHs will merge after
the excitement of eccentricity due to both the KL oscillations and
the BBH-MBH encounters. Note that the contribution from the latter one to the
merging rates is only significant when the binary-single encounters are switched
off. However, as these two effects are less efficient in merging the BBHs
than those of the binary-single encounters, there is a significant drop
in the merging rates for all models (see results in Table~\ref{tab:GW_rate}).

We find that the BBHs in our simulations are all affected by
the dynamical effects mentioned above before they merged. We do not find any
BBHs end up with merger by evolving isolatedly\footnote{Hereafter ``merger by
evolving isolatedly'' means that the merging of BBH is caused only by the GW orbital
radiation in the simulation, and that the BBH does not experience any other dynamical
process, including the binary-single encounters, binary-MBH encounters or KL oscillations,
etc, before the merging event happens.}
for all models except M3c. In model M3c, the merging rate of BBHs evolved isolatedly is
$\sim 0.5$\,Gyr$^{-1}$.  Thus, the merging rates in
Table~\ref{tab:GW_rate} are unique results for BBHs in Galactic
nuclei and are not affected by BBHs evolved in isolation.


\subsection{The eccentricity of the BBH mergers in the LIGO band}

\begin{figure*}
\center
\includegraphics[scale=0.8]{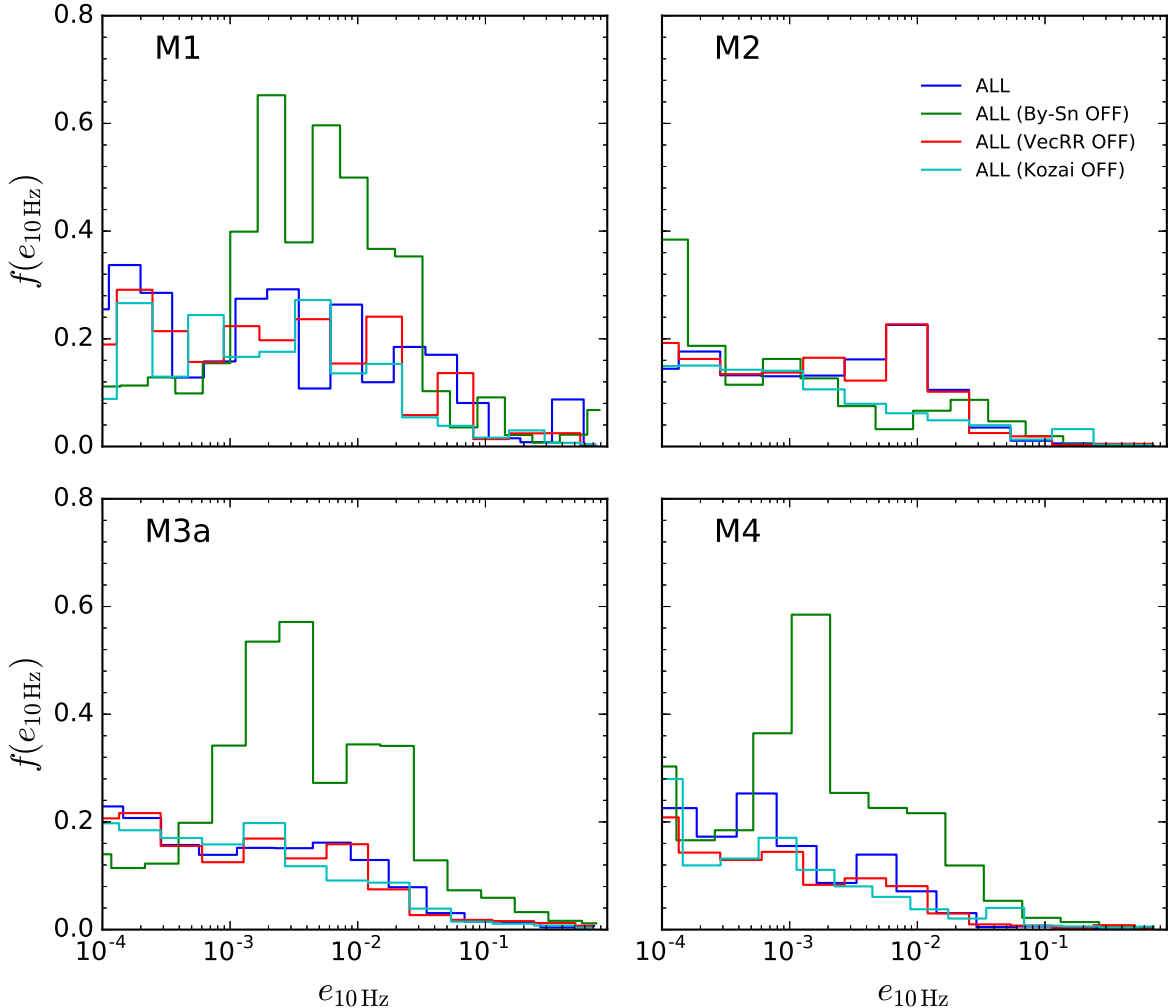}
\caption{The distribution of eccentricity of BBHs when its peak gravitational
  wave frequency enters into the LIGO band, i.e., $f_{\rm GW}=10$ Hz. The solid
  lines show the PDF.
  The blue lines are the results when all the
  dynamical effects are included, while the green, red, and cyan lines are the
  results when binary-single encounter, vector RR or KL effects
only, is turned off, respectively.}
\label{fig:fgw}
\end{figure*}

The BBH remains some eccentricity when the frequency of its
gravitational wave enters into the LIGO band, i.e., $f_{\rm GW}=10$\,Hz. We denote
the eccentricity at this moment as $e_{\rm 10\,Hz}$.
For BBHs in the galactic field, their orbital eccentricities are very
  close to zero when entering the LIGO band \citep{Peters64}. If a BBH has a significant
residual eccentricity, say $e_{\rm 10\,Hz} \gtrsim 0.1$, the modulation in the
waveform will be easily recognized by the matched-filtering techniques
\citep{hinderer17}. In contrast, as we
will see below, the BBHs from galatic nuclei have a non-negligible possibility
to have a significant $e_{\rm 10\,Hz}$.
This will be a smoking-gun effect to
distinguish BBH mergers from galactic fields and nucleus clusters when
enough events are detected to perform a statistical study.

\begin{table}
  \caption{Fraction of eccentric BBHs in nucleus clusters when $f_{\rm
  GW}=10$Hz}
\centering
\begin{tabular}{lccccccccccc}\hline
 model  & ALL        & Vec RR off   & KL off & By-Sn off \\
\hline
M1    & $10.3^a(1.7^b)$ & $12.1(2.5)$   & $7.3(1.8)$   &  $21.6(4.9)$   &    \\
M2    & $4.2(0.46)$     & $4.3(0.79)$
& $4.1(1.5)$   & $5.5(0.76)$    &  \\
M3a   & $5.5(1.1)$      & $4.4(1.4)$   & $4.6(1.0)$   &  $18.5(4.1)$   &    \\
M3b   & $2.8(1.2)$      & $2.5(0.5)$   & $1.1(0.4)$   &  $8.8(3.3)$    &    \\
M3c   & $3.7(0.7)$      & $2.6(0.4)$   & $3.8(0.7)$   &  $14(4.4)$   &    \\
M4    & $1.4(0.3)$      & $1.7(0.6)$   & $2.8(0.6)$   &  $7.7(1.2)$    &   \\
\hline
%
\end{tabular}
\tablecomments{
  $^a$ Percentage of events with $e_{\rm 10\,Hz}>0.01$\\
  $^b$ Percentage of events with $e_{\rm 10\,Hz}>0.05$.
}
\label{tab:GW_ecc}
\end{table}

As we mentioned in earlier sections, the excitement of the eccentricity of BBHs
is mainly through binary-single encounters, BBH-MBH multiple encounters, or KL
oscillations. Different channels excite the eccentricity of the BBHs
in different ways and to different degrees, and thus
the distribution of $e_{\rm 10\,Hz}$ depends
on each or combinations of the above three effects: (1) Binary-single encounters
excite $e_1$ dramatically if $m_\star\sim m_{\rm BBH}$, and slowly if
$m_\star\ll m_{\rm BBH}$. When this effect is switched on,
the excitement of $e_1$ due to BBH-MBH encounters mentioned below will be significantly
suppressed. (2) BBH-MBH encounters excite $e_1$ dramatically
if the encounter is close, e.g., $r_p\la r_t$, and smoothly if $r_p\ga
r_t$~\citep{Zhang10}. Models with massive field stars have large perturbations
in the outer orbits of BBHs and the BBHs can move fast in the lose cone region.
Thus, for these models their excitement of $e_1$ can be dramatic.  (3) The KL
effect can excite $e_1$ to very high values, however, only if some
preferred orbital configurations are satisfied (especially for the inclination
angle). In most cases, the KL
effect changes $e_1$ very smoothly.

Figure~\ref{fig:fgw} shows their probability distribution function
(PDF) of $e_{\rm 10\,Hz}$ when all dynamical effects are
considered or without some of the specific ones.  Table~\ref{tab:GW_ecc}
shows the fraction of BBHs merged with $e_{\rm 10\,Hz}>0.01$ (or $e_{\rm
10\,Hz}>0.05)$ in different models. When all
the dynamical effects are included, we find that there are $\sim2-15\%$
($0.3-5\%$) of the BBHs have $e_{\rm 10\,Hz}>0.01$ ($e_{\rm 10\,Hz}>0.05$).  M1
has the highest probability as in this model the binary-single
encounters excite $e_1$ very dramatically. Note that KL effects
can affect the $e_1$ of BBHs, however, it is not the most possible final
merging channel in all models. Thus, switching it off does not
affect much of the PDF of $e_{\rm 10\,Hz}$. We also do not find
significant difference by switching vector RR off in the simulation.

If the effect of binary-single encounter is switched off, we can see that for
all models (except M2), the $e_{\rm 10\,Hz}$ is higher than the case when the binary-single
encounter is switched on.  For these models (except M2), the BBH-MBH multiple encounters
can excite the eccentricity of BBHs very dramatically, such that $e_1$ can be
very high. For model M2, as $m_\star\ll m_{\rm BBH}$, the outer orbits of
BBHs are perturbed very smoothly, such that they move in and out of the KL
region very frequently ($3r_t<r_p<{\rm min}[20, 0.58(T_{\rm GR}/P2)^{2/3}]r_t$).
 In some rare cases the KL
excites the $e_1$ of BBHs to very high values and thus they merge
quickly. However, in most cases the BBHs move out of the KL region
and their eccentricity are changed only to median values. They then evolve
isolatedly and merge before it enters the KL region again, thus
the overall $e_{\rm 10\,Hz}$ remains low.

\section{The merging events of BBHs in the local universe}

\label{sec:event_rate}

\begin{table*}
\caption{Models}
\centering
\begin{tabular}{lccccccccccccc}\hline
Model & MF$^{a}$ & $r_i^{b}$& $m_\star$ ($\msun$) & $\log$ $\bh$ &
$\langle m_{\rm BBH}\rangle$ & $P(e_{\rm 10\,Hz})^{c}$ & $\mathscr{R}^d$
& $R_{\rm tot}^{e}$ & $R_{\rm tot}^f$ & $P_{\rm tot}(e_{\rm 10\,Hz})^{g}$ \\
\hline
MPW10-5       & PW      &$r_h$& $10$  & $5$ & $21$ & $17 (1.5)$    & $1.43$   & \multirow{5}{*}{$1.6$}  & \multirow{5}{*}{$0.8$} & \multirow{5}{*}{$4.3(0.7)$}  \\
MPW10-6       & PW      &$r_h$& $10$  & $6$ & $21$ & $8.6(0.8)$   & $0.24$   &        &         \\
MPW10-7       & PW      &$r_h$& $10$  & $7$ & $21$ & $2.5(0.9)$   & $0.06$   &        &         \\
MPW10-8       & PW      &$r_h$& $10$  & $8$ & $20$ & $0.5(0.2)$   & $0.03$   &        &         \\
MPW10-9       & PW      &$r_h$& $10$  & $9$ & $19$ & $0.1(0.0)$   & $0.01$   &        &         \\
\hline
MUN10-5       & UN      &$r_h$& $10$  & $5$ & $41$ & $10.(1.4)$  & $3.48$   & \multirow{5}{*}{$3.2$}  & \multirow{5}{*}{$0.9 $} & \multirow{5}{*}{$3.5(0.7)$} \\
MUN10-6       & UN      &$r_h$& $10$  & $6$ & $45$ & $5.4(0.7)$   & $0.84$   &        &         \\
MUN10-7       & UN      &$r_h$& $10$  & $7$ & $43$ & $1.9(0.7)$   & $0.12$   &        &         \\
MUN10-8       & UN      &$r_h$& $10$  & $8$ & $40$ & $0.7(0.3)$   & $0.04$   &        &         \\
MUN10-9       & UN      &$r_h$& $10$  & $9$ & $38$ & $0.3(0.1)$   & $0.01$   &        &         \\
\hline
MUR10-5     & UN &$0.1r_h$      & $10$  & $5$ & $41$ & $17.(1.5)$  &  $24.6$   & \multirow{5}{*}{$20.0$}  & \multirow{5}{*}{$5.6$} & \multirow{5}{*}{$7.5(0.8)$} \\
MUR10-6     & UN &$0.1r_h$      & $10$  & $6$ & $39$ & $10.(0.9)$   & $4.61$   &        &         \\
MUR10-7     & UN &$0.1r_h$      & $10$  & $7$ & $39$ & $5.0(0.5)$   & $0.94$   &        &         \\
MUR10-8     & UN &$0.1r_h$      & $10$  & $8$ & $36$ & $1.2(0.4)$   & $0.19$   &        &         \\
MUR10-9     & UN &$0.1r_h$      & $10$  & $9$ & $36$ & $0.3(0.1)$   & $0.06$   &        &         \\
\hline
MUN1-5        & UN      &$r_h$& $1$   & $5$ & $45$ & $26 (1.6)$    & $21.5$   & \multirow{5}{*}{$30.8$}  & \multirow{5}{*}{$3.1 $} & \multirow{5}{*}{$12(1.3)$} \\
MUN1-6        & UN      &$r_h$& $1$   & $6$ & $35$ & $13 (1.2)$    & $3.77$   &        &         \\
MUN1-7        & UN      &$r_h$& $1$   & $7$ & $36$ & $6.7(1.6)$   & $0.50$   &        &         \\
MUN1-8        & UN      &$r_h$& $1$   & $8$ & $42$ & $2.9(0.8)$   & $0.08$   &        &         \\
MUN1-9        & UN      &$r_h$& $1$   & $9$ & $37$ & $1.8(0.6)$   & $0.01$   &        &         \\
\hline
%
\end{tabular}
\tablecomments{
$^a$ The assumed mass function of BBHs, where ``PW'' means power law
distribution of primary mass that follows $f(m_A)\propto m_A^{-2.35}$, ``UN''
means logarithmic-uniform distribution of primary mass that follows
$f(m_A)\propto m_A^{-1}$.  The mass functions of the secondary component of
these two both follow $f(m_B)\propto m_B^{-1}$ \citep{Abbott16d}.\\
 $^b$ The initial SMA of the outer orbits of BBHs in the cluster. \\
$^c$ Percentage of BBHs with $e_{\rm 10\,Hz}>0.01$ or $e_{\rm 10\,Hz}>0.05$ (in the brackets) in each model.\\
$^d$ Normalized merging rate (Gyr$^{-1}$) in a single galaxy. \\
$^e$ Merging rate (Gpc$^{-3}$ yr$^{-1}$) given by Equation~\ref{eq:totR}, assuming a
constant number fraction of BBHs, i.e., $f_{nb}={\rm min}(10^{-3},10^{-2}
m_\star/\langle m_{\rm BBH}\rangle)$.\\
$^f$ Merging rate (Gpc$^{-3}$ yr$^{-1}$) given by Equation~\ref{eq:totR}, assuming a
constant mass fraction of BBHs, i.e., $f_{mb}=10^{-3}$.\\
$^g$ Percentage of BBHs with $e_{\rm 10\,Hz}>0.01$ or $e_{\rm 10\,Hz}>0.05$ (in the brackets) after all
galaxies are summed.
}
\label{tab:cmodels}
\end{table*}

\begin{figure}
\center
\includegraphics[scale=0.75]{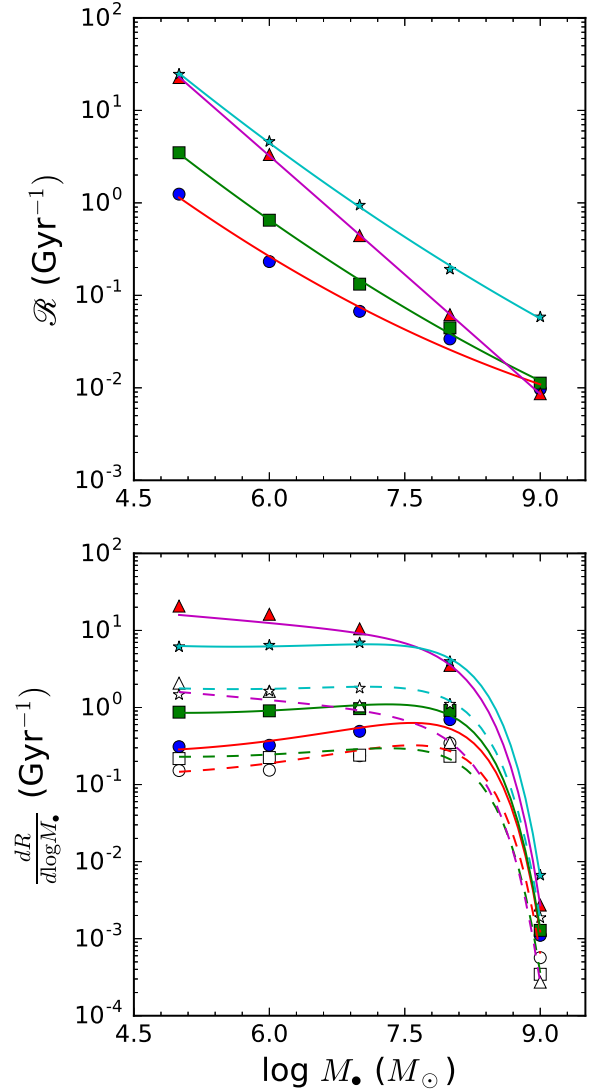}
\caption{Top panel: The normalized event rate $\mathscr{R}$ for galaxies
  containing a MBH with mass $\bh$. Bottom panel: The differential
  merging rate as a function of the MBH mass. We assume
  $\alpha_\star=7/4$. The blue, green, magenta and cyan symbols are the numerical
  results for models MPW10 series, MUN10 series, MUN1 series and MUR10 series,
  respectively. The color lines are the fitting results. In the bottom panel, the
  solid lines are for models assuming $f_{nb}={\rm min}(10^{-3},10^{-2}m_\star/\langle
  m_{\rm BBH}\rangle)$ while the dashed lines are for models assuming
  $f_{mb}=10^{-3}$.
}
\label{fig:cosrate}
\end{figure}

\begin{figure*}
\center
\includegraphics[scale=0.4]{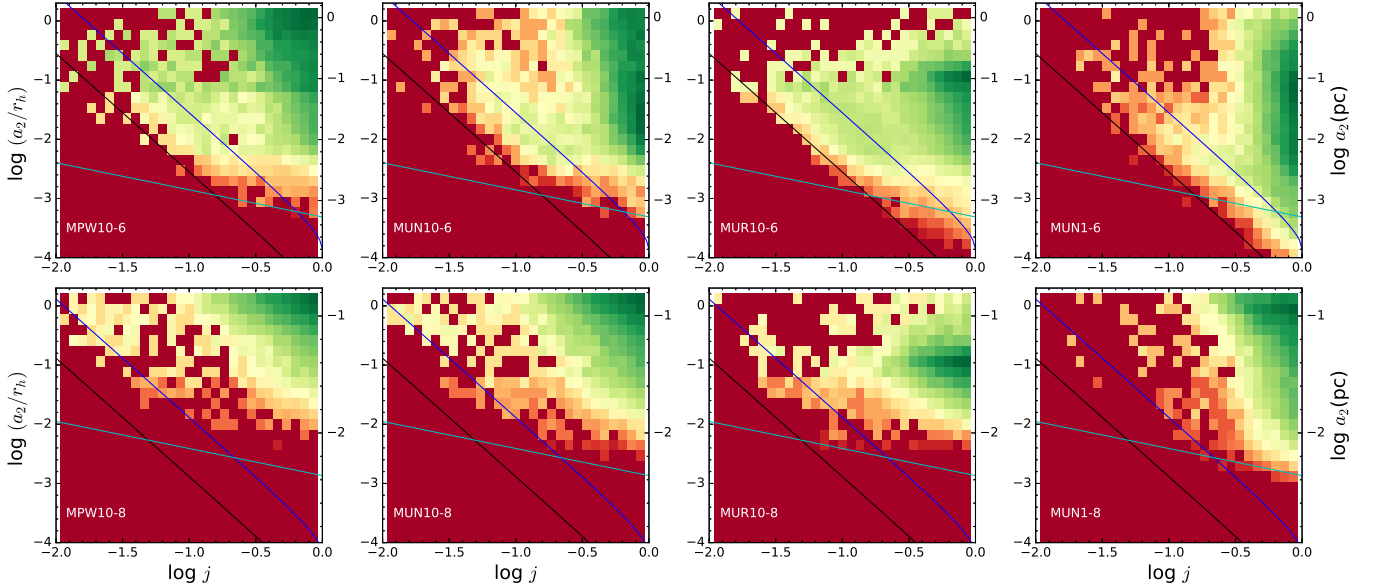}
\caption{The density distribution of $a_2$--$j$ where the BBHs merged in
  different models.  Here $j$ is the dimensionless angular momentum
  $j=\sqrt{1-e_2^2}$.  The blue (black) lines in each panel show the tidal
  radius for a BBH with $a_2=0.5 \, \AU$ ($a_2=0.05 \, \AU$) and $m_A=m_B=5\msun$. The
  green (red) color mesh means high (low) number density regions.
  The cyan solid lines in all panels show the locus of SB.
}
\label{fig:cea}
\end{figure*}

\begin{figure*}
\center
\includegraphics[scale=0.75]{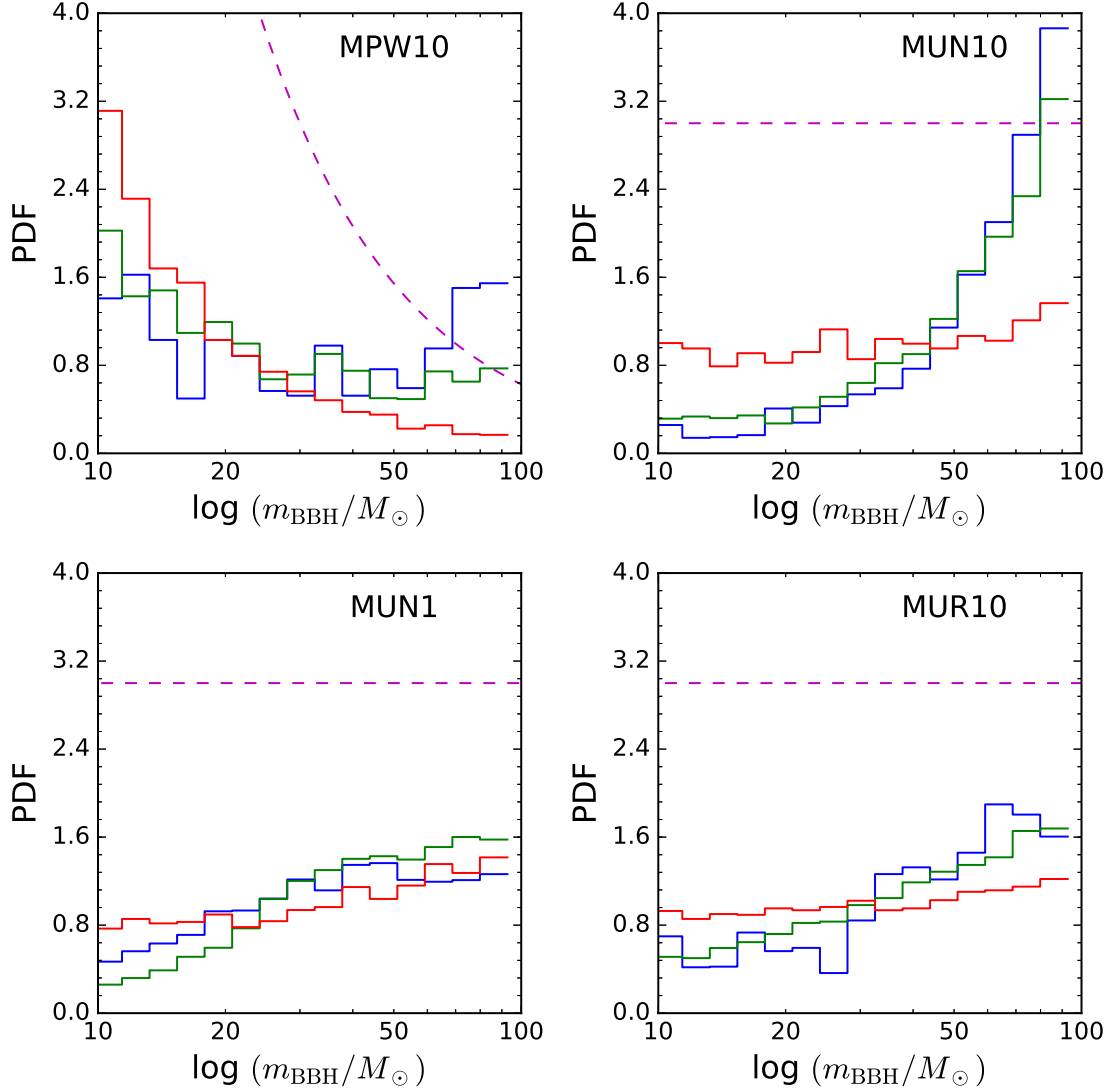}
\caption{Distribution functions of the total mass of the merged BBHs in
  different models.  The magenta dashed lines show the initial mass
  distribution of BBHs. Blue, green and red solid lines show the results for
  galaxies with MBH of mass $10^5\msun$, $10^7\msun$ and $10^9\msun$,
  respectively.
}
\label{fig:mf}
\end{figure*}

\begin{figure*}
\center
\includegraphics[scale=0.75]{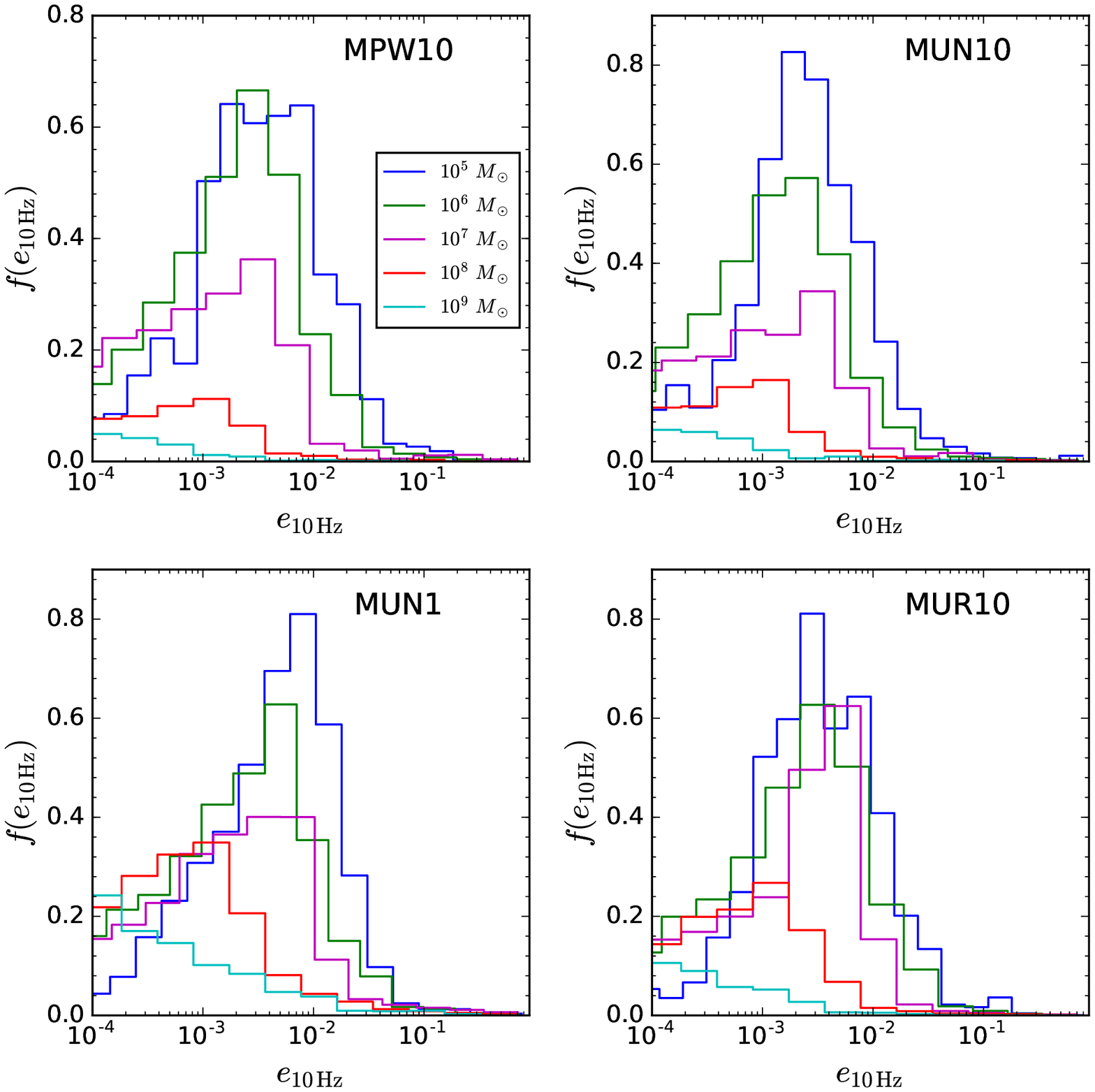}
\caption{The distribution of eccenctricity of BBHs when its peak GW frequency
  enters the LIGO band, i.e., $f_{\rm GW}=10$\,Hz.
  Different panels show results for different models in
  Table~\ref{tab:cmodels}.
  The solid lines in different colors in each panel
  show the results when the mass of MBH in the cluster is different (see
  the legend in the left panel).
}
\label{fig:fgwlu}
\end{figure*}

The event rates shown in Table~\ref{tab:GW_rate} are estimated only for
Milky-Way-like galaxies, and are unrealistic due to their
over simplification of the initial conditions of the BBH populations.  In this
section, we calculate a more realistic estimation of the event rates
basing on population synthesis.

The observed GW rate of merging BBHs is about $12$--$213$\,Gpc$^{-3}$
yr$^{-1}$~\citep{Abbott16d, Abbott17a}. To compare our simulation results with
the observations, we assume two possible cases of the mass distribution
in~\citet{Abbott16d}: (1) a logarithmic-uniform distribution (hereafter the
``UN'' model), i.e., $f(m_A,m_B)\propto m_A^{-1}m_B^{-1}$; (2) a power-law
distribution of primary mass and a uniform distribution on the secondary
mass (hereafter the ``PW'' model), i.e., $f(m_A)\propto m_A^{-2.35}$,
$f(m_B)\propto m_B^{-1}$. In both distributions we
require $5\msun<m_B\le m_A$ and $m_A+m_B<100\msun$.

In previous sections we have assumed that the BBHs are migrated into the
nucleus cluster from the star forming regions
outside the cluster. However, it is also possible
that the BBHs are originated from the star formation processes
within the cluster. For example, our Milky-Way center has an
intense star formation process within the inner parsec~\citep[e.g.,][]{Figer04}.
If so, the KL oscillations may be efficient in
merging these BBHs. To cover these complexities, we explore cases where
initially the BBHs are located at $r_i=r_h$ or $r_i=0.1r_h$.

We assume that the initial inner orbital period distribution of the BBHs
is given by Figure 2 of \citet{Belczynski04}. The period ranges
from $1$ to $10^6$ days.  To reduce the signals of GW merging
rates of the BBHs that have evolved in isolation, without any
impacts from dynamical effects in the galactic nucleus, we
require that initially the GW orbital decay timescale of each
BBH is larger than $1000$\,Myr, i.e., $T_{\rm GW}>1000$\,Myr. Thus, our
estimation of the event rates in this section is likely a conservative one.

The explored four different models are shown in Table~\ref{tab:cmodels}.
For simplicity, we assume $\alpha_\star=7/4$ for these models. Using the method
discribed in Section~\ref{sec:method}, we perform numerical simulations
and obtain the merge event rates for each model for MBH with mass ranging
from $10^5\msun$ to $10^9\msun$.

\subsection{The merging event rates of BBHs}

The results of normalized merging rates for single galaxies are
shown in Table~\ref{tab:cmodels}. We can see that $\mathscr{R}$ is a decreasing
function of $\bh$, and $m_\star$ (see also the top panel of
Figure~\ref{fig:cosrate}).  This is mainly because the larger the mass of the
central MBH, and the mass of the field stars, the softer the BBHs
becomes, and thus the easier for them to be disrupted due to binary-single
encounters. On the other hand, the tidal radius increases with the MBH
mass, thus, the BBHs will be easier to be disrupted around massive MBHs.
Figure~\ref{fig:cea} shows the distribution of the position where the
BBHs merged. We can see that for massive MBH, the BBH only merges at the
outskirt of the nuclei, but for a MBH with a smaller mass, the BBH can merge in
the inner parts of the cluster.  Given the same MBH, if the mass of field stars
is smaller, the BBHs can reach to deeper regions of the cluster.

To estimate the total event rate in nearby universe, we need
to integrate the merge events over all galaxies. The number density of MBHs in
the universe is given by~\citep{Aller02}
\be
\frac{dn}{d\bh}=c_0\left(\frac{\bh}{m_0}\right)^{-1.25}\exp\left(-\frac{\bh}{m_0}\right),
\ee
where $c_0=3.2\times10^{-11} \, \msun^{-1}$ Mpc$^{-3}$ and $m_0=1.3\times10^8\,\msun$.
The cosmological event rate is then
\be
R_{\rm tot}=\int_{10^5\msun}^{10^9\msun} R(\bh) \frac{dn}{d\bh}d\bh
\label{eq:totR}
\ee
Here $R$ is given by Equation~\ref{eq:raten} or~\ref{eq:ratem}, depending on
the model assumption, and we have assumed negligible merging events from
a nucleus cluster with $\bh > 10^9 \msun$ (see Figure \ref{fig:cosrate}).

To obtain the total event rates by Equation~\ref{eq:totR}, we
need to expand the results of $\mathscr{R}$ shown in Table~\ref{tab:cmodels} to
arbitrary MBH masses from $10^5$ to $10^9 \, \msun$.
For MPW10 and MUN10 model series, we have
\be\ba
\log \left( \mathscr{R}/{\rm Gyr}^{-1} \right) \simeq& -1.14 -0.51 \log M_7
+0.04 (\log M_7)^2, \\
\ea\ee
and
\be\ba
\log \left( \mathscr{R}/{\rm Gyr}^{-1} \right) \simeq& -0.83  -0.61 \log M_7
+0.038 (\log M_7)^2,
\ea
\ee
respectively, where $M_7=\bh/10^7\msun$. For MUR10 and MUN1 model series,
we have
\be
\log \left( \mathscr{R}/{\rm Gyr}^{-1} \right) \simeq -0.045  -0.66 \log M_7
+0.030 (\log M_7)^2,
\ee
and
\be
\log \left( \mathscr{R}/{\rm Gyr}^{-1} \right) \simeq -0.34  -0.85 \log M_7
+0.0024 (\log M_7)^2,
\ee
respectively.

The estimated total rates of BBH merging events are shown in the last
two columns of Table~\ref{tab:cmodels}. If assuming a constant number
fraction of BBHs, given $r_i=r_h$ and $m_\star=10 \,\msun$, we find that the total
merging rate is $1.6$ Gyr$^{-1}$ and $3.2$ Gyr$^{-1}$ for models
assuming PW and UN mass functions, respectively. However, the rate can be
up to $\sim 31$ Gyr$^{-1}$ for a UN model and given $m_\star=1 \,\msun$.
Assuming a small $m_\star$ can increase the merging rates as the BBHs become harder
in the cluster and can survive in inner regions of the cluster
after multiple binary-single encounters.
The merging event rates can also be increased to $\sim 20$  Gyr$^{-1}$ for a UN model
but given $r_i=0.1r_h$ and $m_\star=10\,\msun$. As the BBHs are initially located at distance
much closer to the MBH, the KL effects become much more effective in merging these BBHs,
resulting in signficant increase of the merging rates.
In this models, we find that the contribution of merging rates from KL oscillations
is much larger than those from binary-single encounters.

If assuming
a constant mass fraction of BBHs, the rates will be dramatically
reduced.  This is because $\langle m_{\rm BBH}\rangle$ is usually about
$20\msun$--$40\msun$, depending on the assumed mass function (see in
Table~\ref{tab:cmodels}). Such a mass is usually larger (or
much larger) than the mass of the field stars, and reduces
significantly the number of BBHs in each galaxy.  Nevertheless, the event
rate is in orders of $1$--$10$\,Gpc$^{-3}$ yr$^{-1}$, which is still not
negligible for LIGO detections. These results suggest that
the BBH mergers in the center of galaxies can contribute partially to the LIGO observations.

We notice that for MBHs with $10^8-10^9\msun$, the two body relaxation timescale in the cluster
is much longer than the Hubble timescale. Thus, in reality these cluster may never
reach the equilibrium state. Our estimated merging rates for these clusters could be
problematic. Nevertheless, the merging rates contributed from these galaxies are
small (See Figure~\ref{fig:cosrate}), and our estimations of the total merging rates
should not be significantly affected.

\subsection{The mass and eccentricity distribution of the merging BBHs}

Figure~\ref{fig:mf} shows the mass distributions of the merged BBHs in
different models. We can see that, compared with the initial mass distribution
(the dashed magenta lines in each panel), the merged BBHs are likely more
massive. For galaxies containing smaller MBHs with $\bh \la10^7 \,\msun$, the merged
BBHs are also likely more massive than those galaxies containing
larger MBHs  with $ \bh \ga 10^8\,\msun$. More massive BBHs can
survive longer in the galaxies with smaller MBHs, and thus they have
higher event rates for LIGO. Such preference for massive BBHs is less
significant if the BBHs can penetrate into the inner regions of the cluster, as both
the massive and less massive BBHs can be quickly merged in the inner regions, as shown in
the bottom panels of Figure~\ref{fig:mf} (for models MUN1 and MUR10).

Figure~\ref{fig:fgwlu} shows the distribution of the eccentricity of the
BBHs when the GW frequency approaches $f_{\rm GW}=10$\,Hz. We can see
that the eccentricity for low mass MBHs are relatively higher than those around
massive MBHs. For MBH with mass $10^5\,\msun$, we find that $\sim10$--$20\%$ of the
merged BBHs have $e_{\rm 10\,Hz}>0.01$.
Many BBHs are hard in these cluster and they are likely more
compact after each binary-single encounter. Thus, the eccentricity of them in the
merging phase can be high. On the contrary, around massive MBHs the merged
BBHs are commonly less eccentric as many of them are soft binaries, and
that a significant number of them are merging by evolving in isolation.
If the MBH is more massive, i.e., $10^9\,\msun$, the fraction will be about $\la1\%$.
We also find that if the BBHs are initially located at inner regions, e.g.,
$r_i=0.1r_h$ in model MUR10 series, the merging eccentricities can be significantly higher,
as the KL oscillations can be much more effective.

The percentage of BBH mergers with $e_{\rm 10\,Hz}>0.01$ will be around $3$--$10\%$
for all galaxies in the local universe (see the last column of Table~\ref{tab:GW_ecc}).
As comparison, the expected percentage of $e_{\rm 10\,Hz}>10^{-3}$ in globular clusters
is $\sim1\%$ \citep{Rodriguez16b}. For BBH mergers from galactic fields, the final merging
eccentricity is practically zero~\citep{Peters64, Belczynski16}. Thus in principle they can be
distinguished from BBH mergers from the galactic fields and globular clusters using eccentric
waveforms when enough events are accumulated~\citep{hinderer17}.

\section{Discussion}
\label{sec:discussion}

Galactic nucleus cluster is a very complex
system that multiple components of objects, including the stars, white dwarfs,
neutron stars, black holes, and the binaries combined by any of these objects,
are interacting with the central MBH and each other across the evolution history
of galaxies. Although we have included many
dominating dynamical effects, we warn the readers that there
are still a number of complexities that we have not included in our simulation,
which may affect further the merging rate and the dynamical evolutions of the
BBHs.
\begin{itemize}
  \item We have assumed the field stars
    with a single mass, while they should follow a spectrum of mass
    distributions in reality. Due to the mass segregation effect, massive stars/objects
    are concentrated in inner regions while those less massive ones are more
    likely at the outer parts.  Thus, the dynamics of BBHs in the outer parts
    of the cluster may be different from the inner parts. Additionally, we
    do not trace and discuss those BBHs if one of its components
    is exchanged by the field stars, and we do not consider the case
    that the isolated black holes in the cluster can be
    captured into binary systems. These complexities can be
    considered in the future, if we can simulate the relaxation process by
    methods similar to those in~\citet{Henon71} and~\citet{Joshi00}.

  \item In estimating the merging event rates, we have assumed
    a constant mass or number ratio of stellar-mass BBHs
    with respect to the background field stars. In reality, neither of these two cases may be
    true, as the number of BBHs may vary from galaxies to galaxies, and may
    depend on various properties of the nucleus
    cluster and the black holes.

  \item We have assumed that the BBHs formed continuously inside or outside
  of the cluster, as suggested by the continuous star formation history observed
  in the nuclear star cluster in many galaxies~\citep[e.g.,][]{Walcher06}.
  This assumption may have over simplified the complex star formation history of nuclear
  star clusters under the cosmological context, where the infalling gas due to the
  merging of galaxies, migration of stellar clusters and the existence of binary MBHs
  may affect the supply of BBHs~\citep[e.g., ][]{Antonini15,Arcasedda18,Arcaseddda19}.

\end{itemize}

Our treatment of the vector RR process may be oversimplified.
However, even with a more sophisticated treatment, e.g., given
by~\citet{Hamers18}, they find that the vector RR enhances the merger
rate only with MBHs of small masses, and drops sharply with
increasing MBH masses. Thus, we consider that the
details have no significant effects on our results.

 We have not included the precession of the outer orbit due to the distributed mass
when calculating the Kozai-Lidov oscillations in the current simulation. To explore
the impact on our results due to such simplification, we perform additional simulations
that include such precession of the outer orbit. By comparing the simulation results,
we find only a slight reduce on the merging rates from Kozai-Lidov mechanism, and our
main results on the merging rates will not be significantly affected.
This is consistent with those in~\citet{Hoang18}.

Our work can be expanded to consider the evolution and the merging of neutron
star binaries, or the neutron star--black hole binaries. Currently it is still
difficult to distinguish the merging channels of BBHs as their localizations are
challenging, as there are no definite electromagnetic
counterparts observed for these events. However, the
merging of neutron star binaries or neutron star--black hole
binaries in galactic nucleus are supposed to have
electromagnetic counterparts \citep{Abbott17d}
and their localizations can be quite accurate.  If these events can be
detected by LIGO and that its origin can be
confirmed by their
  electromagnetic counterparts or the measurement of the eccentricity, their
  properties can be used further to study the dynamics around the MBHs. We will
  defer such studies to future.

There are a number of important difference between the evolutions of BBHs that
are located in the globular clusters and the galactic nuclei.
For example, most of the BBHs residing in globular clusters are
hard binaries, and every binary-single encounters are likely
to harden them. The hardening of BBHs will usually eject them
out of the globular cluster, and thus their mergers are likely taking place
outside of the globular cluster. However, in a galactic
nucleus, the BBHs are most likely soft binaries, and they are
likely be softened and ionized by the binary-single encounters.  On the other
hand, if they merge, they merge within the galactic nucleus. In a
galactic nucleus, there are unique dynamical processes, for example, the
resonant relaxations, the KL effects and tidal disruptions of BBHs by the central MBHs. Thus, the event rate and statistical properties of their
GW signals will likely be different from those of the globular
clusters.  By comparing their difference, we may likely  distinguish
the GW events in these two scenarios with future observations.

Although we find that our Monte Carlo scheme can reproduce the theoretical
expectations of individual dynamical effects, we still can not guarantee that
the combinations of these effects have well captured the detailed evolutions of the
BBHs around MBHs.  Thus, our numerical method still needs verifications (or
calibrations) by $N$-body simulations that can combine well of the dynamical
effects that we have considered.  Though rewarding, these $N$-body
simulations are very challenging and expensive, which are out of the scope
of this study. We defer it to future investigations.

\section{conclusions}
\label{sec:conclusion}

One of the possible channels to merge the stellar mass binary
black holes (BBHs) is in a galactic nucleus that contains a MBH in the
center. In this work, we study the dynamical evolution of the BBHs in
a galactic nucleus that contains a MBH. Along the pathway of their final merging,
we consider simultaneously the non-resonant and resonant relaxations of the
BBHs, the binary-single encounters of the BBHs with the field stars, the
Kozai-Lidov (KL) oscillation and the close encounters between the BBHs and the
central MBH, which usually lead to BBHs' tidal disruptions. These effects have
been individually studied and discussed in lots of previous work, however, not
yet well combined. Here we consider all of them in a Monte-Carlo scheme to
study the merging of BBHs for a comprehensive study.

We find that the tidal disruption of BBHs by MBHs can effectively reduce the
number of the BBHs in the inner regions of a nucleus cluster,
especially if resonant relaxation (RR) are being considered.  The binary-single
encounters can further reduce the number of BBHs within the cluster, as most of
the BBHs become soft binaries.
Each encounter can possibly increase the separations of the BBHs and
cumulatively lead to the ionization of the BBHs.  In the meanwhile, the
binary-single encounters can also make the binary hardening, and increase the
eccentricity of the BBHs. These effects can increase the merging rates
significantly.

If the total mass of the BBH is heavier than the mass of the field stars,
they can sink to the central regions by mass segregation effects,
making the KL oscillation an important factor to trigger their
mergers. For other dynamical effects, for example, the
vector RR process, we do not see significant difference of the
merging rates with or without it.

We find that in our simulations the mergers of BBHs are mainly
contributed by galaxies containing  MBHs that are less massive than $10^8 \,\msun$, and
the total event rates are likely in orders of $1$--$10$\,Gpc$^3$ yr$^{-1}$,
depending also on the detailed assumptions of the
nucleus clusters. The eccentricity of the BBHs ($e_{\rm 10\,Hz}$)
when its peak GW frequency enters into the observational bands of the Advanced
LIGO/Virgo detectors (e.g., $10$\,Hz)
depends on the initial conditions of the cluster.  It is likely that there are
$3$--$10\%$ of BBHs with $e_{\rm 10\,Hz}>0.01$, and the probability is higher for
clusters harboring MBHs with a smaller mass. This will be a smoking-gun signal
  when solid statistical analysis becomes available with accumulating events.
  Our results show that the mergers of BBHs within galactic nuclei
  can be one of the important sources of the
  merging events detected or to be detected by ground-based GW detectors.

\acknowledgements

\noindent
We thank the anonymous referee for the helpful comments that have
improved this paper.
This work was supported in part by the National Natural Science Foundation of China
under grant No. 11603083, 11673077. This work was also supported in part by
Guangzhou university Startup funds grand No. 69-18ZX10362,
``the Fundamental Research Funds for the Central Universities'' grant No. 161GPY51,
the Key Project of the National Natural Science Foundation of China under grant No.
11733010.
LS was supported by the Young Elite Scientists Sponsorship Program by the China Association for Science and Technology (2018QNRC001), and partially supported by the National Natural
Science Foundation of China (11721303), XDB23010200.
The simulations in this work are performed partly in the TianHe II National
Supercomputer Center in Guangzhou, and partially on the computing cluster in School
of Physics and Astronomy, Sun Yat-Sen University.


\appendix

\section{The dynamics of the outer orbits}
\subsection{The diffusion coefficients in two-body relaxations}
\label{apx:two_body}

The diffusion coefficients $D_{EE}$ and $D_{JJ}$ describe the orbit-averaged scatterings of
the energy and angular momentum of the orbit for point-like particles around the
MBH.  Similarly, $D_{E}$ and $D_J$ describe the orbit-averaged drifts of the energy and
angular momentum, respectively, and $D_{EJ}$ describes the correlations between
the two.  The calculations of the diffusion coefficients have been done and
discussed in many studies~\citep[e.g.,][]{SM78, Cohn78, Baror16}. Here
we use the formalism in~\citet{Baror16} where the mass of the
binary can be different from the field stars. Denote the SMA and the
eccentricity of the particle cycling around the MBH as $a_2$
and $e_2$, and suppose that the dimensionless distribution function of the
field stars is given by $f(E)$.  Denote $m_{\rm BBH}$ and $m_\star$ as the
masses of the binary and the field star respectively. Then firstly the function
$\Gamma_{ijk}$ and $\Gamma_0$ are calculated,
\be\ba
\Gamma_{ijk}&=2^{1+k-i}\frac{\kappa}{\pi}\int^{1}_{-1} dy\int^{2/(1+y)}_{1} ds
f_a(sE)\\
&\times\frac{1}{\sqrt{1-y^2}}\frac{(1+ye_2)^i}{(v^2/E)^k}(v_a/v)^j\\
\Gamma_0&=\kappa \int^1_{-\infty} ds f_a(sE).
\ea\ee
Here $\kappa=(4\pi Gm_\star)^2 \ln \Lambda$, $y=(r/a_2-1)/e$,
$v_a^2=2E(2a_2/r-s)$, $v^2=2E(2a_2/r-1)$, $\Lambda=\bh/m_\star$. Then the
diffusion coefficients are given by
\be\ba
D_E/E&=\frac{m_{\rm BBH}}{m_\star}\Gamma_{110}-\Gamma_0,\\
D_{EE}/E^2&=\frac{4}{3}\Gamma_{13-1}+\Gamma_0,\\
D_{J}/(jJ_c)&=\frac{5-3j^2}{12}\Gamma_0-j^2\frac{m_{\rm BBH}+m_\star}{2m_\star}\Gamma_{111}+\Gamma_{310}-\frac{1}{3}\Gamma_{330},\\
D_{JJ}/J_c^2&=\frac{5-3j^2}{6}\Gamma_0+\frac{j^2}{2}\Gamma_{131}-\frac{j^2}{2}\Gamma_{111}
+2\Gamma_{310}-\frac{2}{3}\Gamma_{330},\\
D_{EJ}/(EJ_c)&=-\frac{2}{3}j(\Gamma_0+\Gamma_{130}).
\ea\ee
Here $j=J/J_c$ is the dimensionless angular momentum, and $J_c=\sqrt{G\bh a_2}$
is the maximum angular momentum.  Note that $j$ could not be either zero nor
one, otherwise the integration will be divergent.

\subsection{The scalar and the vector resonant relaxations}
\label{apx:resonant}

The scalar and vector resonant relaxations change only the angular momentum of
the outer orbit of the binary. For the scalar resonant relaxation, we take
formalisms similar to~\citet{Baror16}. Define $Q=\bh/m_\star$ as the number of
field stars and $\nu_r=2\pi/P(a_2)=a_2^{-3/2}\bh^{1/2}G^{1/2}$ as the Keplerian
orbital frequency, the RR torque $\tau_N$ is given by
\be
\tau_N=0.28\sqrt{1-j}\sqrt{N(2a_2)}Gm_\star/a_2.
\label{eq:tau_N}
\ee
Here $N(a_2)=\bh/m_\star(a_2/r_h)^{3-\alpha}$.
Denote $\nu_J=\tau_N/J_c$, the diffusion coefficients
are given by
\be
D_{JJ}^{\rm RR}=\frac{2J_c^2\nu_J^2(j)T_c(a_2)}{1+[T_c(a_2)\nu_p(a_2,j)]^2}.
\ee
Here
\be
\nu_p=|\nu_M+\nu_{\rm GR}|,
\label{eq:nup}
\ee
is the precession frequency, where
\be
\nu_{\rm GR}=\frac{3r_g}{a_2j^2}\nu_r(a_2)
\label{eq:ngr}
\ee
is the frequency of the GR-induced precession.  $\nu_M$  has an
exact form when $\alpha=2$,
\be
\nu_{\rm M}(\alpha=2, j)=-\frac{N(a_2)}{Q}\frac{j}{j+1}\nu_r.
\label{eq:numalpha2}
\ee
For other values, it can be obtained numerically. We find that
\be
\nu_{\rm M}(\alpha, j)=-\frac{N(a_2)}{Q}\frac{j^{2-\gamma}(1-j^{\gamma})}{1-j^2}\nu_r.
\ee
When $\alpha=7/4$, $\gamma=1.48$; when $\alpha=1/2$, we find that $\gamma=4.0$;
when $\alpha=2$, $\gamma=1$, which reduces to
Equation~\ref{eq:numalpha2};  when $\alpha=1$, we find that $\gamma=3.37$.

$T_c(a_2)$ is given by (note that the coherence time $T_c$ here
has not  included the GR precession~\citep{Baror16})
\be
T_c(a_2)=\sqrt{\pi/2}\nu_M^{-1}(2a_2, \sqrt{1/2}).
\label{eq:tca2}
\ee
The drift term is given by
\be
D_J^{\rm RR}=\frac{1}{2J}\frac{\pl (J D_{JJ}^{\rm RR})}{\pl J}.
\ee

For the vector RR relaxtion, we take a very simple model, where the direction of
the angular momentum is randomly walking on the
sufrace of the sphere, with the unity change happens in a timescale given by
$T_{\rm RR}^v$ in Equation~\ref{eq:trrv}. We generate a random
vector $\hat{dj}$ with magnitude $l=(dt/T_{\rm RR}^v)^{1/2}$,
where $l$ is the standard deviation of a log-normal distribution.
The unit angular momentum of the CM is changed to $\hat j'=\hat j+
\hat{dj}$; the direction of $\hat{dj}$ is set such that $\hat j'$
is always a unit vector.  Here $\hat j$ is the unit angular momentum.  First we
solve a reference vector $\hat{dj_0}=(0,l\cos s, l\sin s)$ where
$|\hat j+ \hat{dj}_0|=1$ (If $j_y^2+j_z^2<l^2/4$, we can set $\hat{dj}_0=(l\cos
s, 0, l\sin s)$).
Then the random vector $\hat{j}'$ is generated after rotating $\hat
j+\hat{dj_0}$ around the vector $\hat j$ by a random angle $\phi\in(0,2\pi)$.
If $l>1$, we simply set the angular momentum randomly distributed.

\section{The GW frequency of the merging BBHs}
\label{apx:GWfreq}

The power of GWs emitted from an eccentric in-spiraling BBH covers a
broad range of frequency and the maximum occurs at the peak frequency $f_{\rm
GW}$, which is given by~\citep{Wen03}
\be
f_{\rm GW}=\frac{\sqrt{(m_A+m_B)G}}{\pi}\frac{(1+e_1)^{1.1954}}{a_1^{3/2}(1-e_1^2)^{3/2}}.
\label{eq:fgwob}
\ee
If the BBHs are not affected by the KL effects, during their
merging process, the evolution of $a_1$ and $e_1$ are given
by~\citep{Peters64}
\be
a_1(e_1)=\frac{c_0 e_1^{12/19}}{1-e_1^2}\left(1+\frac{121}{304}e_1^2\right)^{870/2299}.
\label{eq:gwae}
\ee
Here $c_0$ is a constant determined by the initial parameters.

In our simulation, we mainly focus on the eccentricity of the BBHs when $f_{\rm
GW}=10$\,Hz, i.e., $e_{\rm 10\,Hz}$. When a BBH can be considered as a merger in the
simulation (see Section~\ref{subsec:MC_schemes}), $e_{\rm 10\,Hz}$ is
calculated according to the details of the simulation: (1) During the merge, if
the BBHs are not affected by the KL effects, we obtain the value of $e_{\rm
10\,Hz}$ by setting $f_{\rm GW}=10$\,Hz in
Equation~\ref{eq:fgwob} and combining with
Equation~\ref{eq:gwae}.  The constant $c_0$ is determined according
to the last value of $e_1$ and $a_1$ of the BBHs.  (2) If the BBHs are affected
currently by KL effects and that KL term dominates over the GW term, i.e.,
$|\dot e_1^{\rm KL}|>|\dot e_1^{\rm GW}|$, we then record the value of $e_1$ as
$e_{\rm 10\,Hz}$ when $f_{\rm GW}=10$\,Hz. (3) If the BBHs are affected currently
by KL effects and that GW term dominates over the KL term, i.e.,  $|\dot
e_1^{\rm KL}|<|\dot e_1^{\rm GW}|$, we calculate the value of $e_{\rm 10\,Hz}$
similar to (1).

\section{The binary-single encounters}
\label{apx:encounter}

Here we first calculate the rates of binary-single encounters in a
nucleus cluster.  Suppose that the density profile of the field
stars is stable, then the rate of binary-single encounters is given
by~\citep{Sigurdsson93}
\be
R_{\rm EC}(a_2)=\langle n(r)\Sigma v_\infty\rangle =\frac{2}{P}\int
\int^{r_a}_{r_p} n(r,v') \Sigma v_\infty (a_2) \frac{dr}{v_r} d^3v',
\ee
where $\Sigma$ is the cross section of binary-single encounter.  $v_\infty=
|\vec{v}-\vec{v}'|$ is the relative velocity between the binary and the star.
$v=2(E+G\bh/r)$, $v_r$, $r$ and $a_2$ is the velocity, radial velocity,
position and outer SMA of the binary respectively. $r_a=a_2(1+e_2)$ and $r_p=a_2(1-e_2)$ is
apocenter and pericenter of the outer orbit,
respectively.

The cross section for percienter passage less than $p$ is given by
\be
\Sigma=\pi p^2\left(1+\frac{2m_TG}{p v_\infty^2 }\right).
\ee

Assuming that the velocity is isotropic, then we have $n(v',r)=n(v')n(r)$,
where $n(r)=n_0(r/r_h)^{-\alpha_\star}$ and $n(v')$ satisfies a
Maxiwell velocity distribution, with the velocity dispersion $\sigma_a$
satisfying~\citep{Alexander05}
\be
\sigma_a^2=\frac{1}{1+\alpha_\star}\frac{G\bh}{r}.
\ee

The integration on $v'$ can be preformed independently
\be\ba
\langle v_\infty\rangle&=\int n(v')v_\infty d^3v'=
\frac{2\sigma_a}{(2\pi)^{1/2}}
\exp\left(-\frac{{v}^2}{2{\sigma}_a^2}\right)
+\left(v+\frac{\sigma_a^2}{v}\right){\rm erf}\left(\frac{v}{\sqrt{2}\sigma_a}\right),
\ea\ee
and~\citep{Binney87}
\be\ba
\left\langle \frac{1}{v_\infty}\right\rangle&=
\int \frac{n(v')}{v_\infty} d^3v'=
\frac{1}{v}{\rm erf}\left(\frac{v}{\sqrt{2}\sigma_a}\right).
\ea\ee
We denote
\be\ba
\Theta&=\frac{2}{P}
\int^{r_a}_{r_p}   \left(\frac{r}{a_2}\right)^{-\alpha_\star}
\frac{\langle v_\infty\rangle}
{\sigma_a}\frac{dr}{v_r},
\ea\ee
and
\be\ba
\Phi&=\frac{2}{P}
\int^{r_a}_{r_p}   \left(\frac{r}{a_2}\right)^{-\alpha_\star}
\sigma_a\left\langle\frac{1}{v_\infty}\right\rangle\frac{dr}{v_r}.
\ea\ee
Here $\Theta$ and $\Phi$ are functions of $e_2$ $\alpha_\star$
only. By numerical simulations, when $\alpha_\star=7/4$,
we find that $\Theta\simeq 1.36+0.49e_2^2$ and $\Phi\simeq 0.86+2.5e_2^2$; when $\alpha_\star=1$,
we have $\Theta\simeq 1.47$, $\Phi\simeq 0.84+1.1e_2^2$.

Then $R_{\rm EC}$ can be rewritten as
\be\ba
R_{\rm EC}(a_2)=&
\pi p^2 n_0\sigma_a
\left(\frac{a_2}{r_h}\right)^{-\alpha_\star}
\Theta
+\frac{2\pi p m_T G n_0}{\sigma_a }
\left(\frac{a_2}{r_h}\right)^{-\alpha_\star}
\Phi.
\label{eq:rbi}
\ea\ee

Then the probability of taking $k$ times of binary-single encounters
in time interval $\delta t$ is given by the Poisson distribution
\be
P_{\rm EC}(k)=e^{-\lambda}\frac{\lambda^k}{k!},
\label{eq:pec}
\ee
where $\lambda=R_{\rm EC}\delta t$.

Occasionally, the binary will encounter with the single
star with distance less than $0.01\, \AU$. In these cases, the binary will
experience an extremely strong encounter.  However, if the incoming
single object is a star, then the black hole touches, almost, to the surface of
the star, leading to modifications of the trajectory due to tidal effects.  If
the incouming object is a black hole, then the gravitational wave can lead to
the decay of the orbital energy, too. To avoid the complexities of these cases,
we set a softening radius of $10^{-3}\,\AU$ between the incoming star and the
black holes in the $3$-body calculations.

Currently, we have neglected any relativistic effects on the binary-single encounters
to reduce the computational costs. In principle, if the incoming object is a stellar
mass black hole, we need to additionally consider the post-Newtonian
corrections of the orbit to include high order relativistic effects such as spin
effects or the GW orbital decay. Such simplification may result in under estimation of the merging
event rates. Considering that the fraction of stellar mass black holes in a nuclear cluster
should be very small, e.g., $<10^{-2}$, the probability of such BBH-single black hole
encounters should be small, and thus the impact on the merging event rates should not be
significant. Nevertheless, we can include the post-Newtonian corrections into our 3-body
simulations in future studies.

   The initialization and termination of simulations of the binary-single encounters
are considered as follows: (1) Initially the mass center of BBH
and the single star are separated at a distance of $r_{bi}=50a_1$ away from each other,
where $a_1$ is the inner SMA of the BBH;
(2) The mass center of BBH and the single star form a two
body system. Denote $a'_2=-m_T/v_\infty^2$
as the semimajor axis (SMA) of the orbit of this two body
system and $\mathcal{M}$ as the mean anomaly of the BBH at distance $r_{bi}$,
 where $m_T$ is the total mass of BBH-single star system and $v_\infty$ is the
 relative infinite velocity between the binary and the single star. Then
we perform the 3-body simulation for a duration $dt=2|\mathcal{M} |/(2\pi)(|{a'_2}^3|
/m_T)^{1/2}$. In a two-body problem,
after such duration the mass center of BBHs will take one pericenter passage and return to
the distance of $50a_1$ again; (3) We determine the outcome of the BBH-single encounter.
If we find that it is a fly-by, ionization or exchange event, the simulation stops;
(4) However, if it is not any of the outcome in (3), usually the 3-body either form a triple system,
or the 3-body are still in chaotic orbits, and we continue
the simulations for additional time $dt\rightarrow1.5dt$ until it becomes any results in
(3). We repeat (4) for at most $10$ times such that the total time of simulations can be
$\sim170$ times of the original value of $dt$ in (2). In most cases, the binary will
end up with one of the results listed in (3), but if not, we simply abandon the
event and remove the BBH from the Monte-Carlo simulation.

When there are multiple encounters, i.e., $n_{\rm EC}>1$ in one time step $\delta t$ of our
simulation (See Section~\ref{subsec:MC_schemes} and~\ref{subsec:tstep}), we assume that the first one of them
occur within time of $\delta t/n_{\rm EC}$. After the first encounter, if it's a flyby event,
the SMA and the eccentricity of the binary is changed. In the rest of the time step, i.e.,
$\delta t(n_{\rm EC}-1)/n_{\rm EC}$, we calculate the collision rate and the expected number of
collisions, i.e., $n_{\rm EC}'$, according to the updated orbits of the binary, and perform a
successive encounter if $n_{\rm EC}'\ge1$. This repeats until there
is no more successive encounters in the remain time of $\delta t$.

Currently, we have ignored the correction of the outer orbit due to the binary-single encounter.
The conversion between the energy of the inner and outer orbits during the binary-single encounter should be
in orders of $\Delta E\sim\epsilon$, where $\epsilon$ is the energy of the inner orbit.
Such simplification should not lead to significant differences as $\epsilon$ is usually much
smaller than the energy of the outer orbit of the BBHs. Nevertheless,
we will introduce corrections on both
the energy and angular momentum of the outer orbits due to binary-single encounters in the future.

\end{document}